%% file: main.tex
\documentclass[manuscript,screen]{acmart}

\usepackage{float}
\usepackage[nolist]{acronym}
\usepackage{comment}
\usepackage{verbatim}

\usepackage{array}
\newcolumntype{H}{>{\setbox0=\hbox\bgroup}c<{\egroup}@{}}
\newcolumntype{P}[1]{>{\centering\arraybackslash}p{#1}}
\newcolumntype{L}[1]{>{\RaggedRight\hspace{0pt}}p{#1}}
\newcolumntype{R}[1]{>{\RaggedLeft\hspace{0pt}}p{#1}}
\usepackage{longtable}
\usepackage{multirow}
\usepackage{pbox}
\usepackage{rotating}
\usepackage{makecell}
\usepackage{graphicx}
\usepackage{booktabs}
\usepackage[section]{placeins}

\usepackage{floatflt,epsfig, wrapfig}

\newif\iftitlepage
\titlepagefalse 

\renewcommand{\arraystretch}{1.25} 

\usepackage{pdfrender}

\usepackage{pifont}

\usepackage{listings, xcolor}

\definecolor{verylightgray}{rgb}{.97,.97,.97}

\lstdefinelanguage{Solidity}{
	keywords=[1]{anonymous, assembly, assert, balance, break, call, callcode, case, catch, class, constant, continue, constructor, contract, debugger, default, delegatecall, delete, do, else, emit, event, experimental, export, external, false, finally, for, function, gas, if, implements, import, in, indexed, instanceof, interface, internal, is, length, library, log0, log1, log2, log3, log4, memory, modifier, new, payable, pragma, private, protected, public, pure, push, require, return, returns, revert, selfdestruct, send, solidity, storage, struct, suicide, super, switch, then, this, throw, transfer, true, try, typeof, using, value, view, while, with, addmod, ecrecover, keccak256, mulmod, ripemd160, sha256, sha3}, 
	keywordstyle=[1]\color{blue}\bfseries,
	keywords=[2]{address, bool, byte, bytes, bytes1, bytes2, bytes3, bytes4, bytes5, bytes6, bytes7, bytes8, bytes9, bytes10, bytes11, bytes12, bytes13, bytes14, bytes15, bytes16, bytes17, bytes18, bytes19, bytes20, bytes21, bytes22, bytes23, bytes24, bytes25, bytes26, bytes27, bytes28, bytes29, bytes30, bytes31, bytes32, enum, int, int8, int16, int24, int32, int40, int48, int56, int64, int72, int80, int88, int96, int104, int112, int120, int128, int136, int144, int152, int160, int168, int176, int184, int192, int200, int208, int216, int224, int232, int240, int248, int256, mapping, string, uint, uint8, uint16, uint24, uint32, uint40, uint48, uint56, uint64, uint72, uint80, uint88, uint96, uint104, uint112, uint120, uint128, uint136, uint144, uint152, uint160, uint168, uint176, uint184, uint192, uint200, uint208, uint216, uint224, uint232, uint240, uint248, uint256, var, void, ether, finney, szabo, wei, days, hours, minutes, seconds, weeks, years},	
	keywordstyle=[2]\color{teal}\bfseries,
	keywords=[3]{block, blockhash, coinbase, difficulty, gaslimit, number, timestamp, msg, data, gas, sender, sig, value, now, tx, gasprice, origin},	
	keywordstyle=[3]\color{violet}\bfseries,
	identifierstyle=\color{black},
	sensitive=false,
	comment=[l]{//},
	morecomment=[s]{/*}{*/},
	commentstyle=\color{gray}\ttfamily,
	stringstyle=\color{red}\ttfamily,
	morestring=[b]',
	morestring=[b]"
}

\usepackage{hyperref}

\AtBeginDocument{%
  \providecommand\BibTeX{{%
    \normalfont B\kern-0.5em{\scshape i\kern-0.25em b}\kern-0.8em\TeX}}}

\setcopyright{acmcopyright}

\begin{document}

\input{00-Abbreviations}

\title[E-prescription management through blockchain and digital wallets]{Harmonizing sensitive data exchange and double-spending prevention through blockchain and digital wallets: The case of e-prescription management}

\author{Vincent Schlatt}
\email{vincent.schlatt@fim-rc.de}
\orcid{0000-0002-3856-9302}
\author{Johannes Sedlmeir}
\email{johannes.sedlmeir@fim-rc.de}
\orcid{0000-0003-2631-8749}
\affiliation{%
  \institution{FIM Research Center, University of Bayreuth}
  \country{Germany}
}
\author{Janina Traue}
\email{janina.traue@uni-bayreuth.de}
\orcid{0000-0002-2655-3848}
\affiliation{%
  \institution{University of Bayreuth}
  \country{Germany}}

\author{Fabiane V{\"o}lter}
\orcid{}
\email{fabiane.voelter@fit.fraunhofer.de}
\orcid{0000-0002-4174-444X}
\affiliation{%
  \institution{Branch Business \& Information Systems Engineering of the Fraunhofer FIT}
  \country{Germany}
}


\input{00-Abstract}



\maketitle

\clearpage

\input{01-Introduction}
\input{02-Background}

\input{03-Method}
\input{04-Requirements}
\input{05-Architecture}

\input{06-Evaluation}
\input{07-Discussion}
\input{08-Conclusion}

\begin{acks}
We gratefully acknowledge the Bavarian Ministry of Economic Affairs, Regional Development and Energy for their funding of the project ``Fraunhofer Blockchain Center (20-3066-2-6-14)'' that made this paper possible. We also want to express our special thanks to Matthias~Babel for his support with the e-prescription smart contract, to Jana~Gl\"ockler for her groundwork on connecting the Hyperledger Aries Cloud Agent in Python with the Django framework, and to Jonathan~Lautenschlager and Felix~Paetzold for their valuable feedback and suggestions for improvement. Finally, we want to thank the editor and the anonymous reviewers for their valuable guidance and recommendations for improvement during peer review.
\end{acks}

\clearpage
\bibliographystyle{ACM-Reference-Format}
\bibliography{90-Literature}

\clearpage
\appendix
\section*{Appendix}
\input{91-Appendix}

\end{document}

%% file: 00-Abbreviations.tex
\begin{acronym}[e-prescription]
\setlength{\itemsep}{-0.25cm}

\acro{AWS}[AWS]{Amazon web services}
\acro{BFT}[BFT]{byzantine fault tolerant}
\acro{CA}[CA]{certificate authority}
\acro{CCPA}[CCPA]{California consumer privacy act}
\acro{DID}[DID]{decentralized identifier}
\acro{DLT}[DLT]{distributed ledger technology}
\acro{DLPS}[DLPS]{distributed ledger performance scan}
\acro{DP}[DP]{design principle}
\acro{DPKI}[DPKI]{distributed \ac{PKI}}
\acro{DSR}[DSR]{design science research}
\acro{E2E}[E2E]{end-to-end}
\acro{EBSI}[EBSI]{European blockchain service infrastructure}
\acro{eIDAS}[eIDAS]{electronic Identification, Authentication and Trust Services}
\acro{EU}[EU]{European Union}
\acro{e-prescription}[e-prescription]{electronic prescription}
\acro{e-health}[e-health]{electronic health}
\acro{GDPR}[GDPR]{general data protection regulation}
\acro{HIPAA}[HIPAA]{health insurance portability and accountability act}
\acro{HL7CDA}[HL7~CDA]{Health~Level~7~Clinical Document Architecture}
\acro{IBFT}[IBFT]{Istanbul byzantine fault tolerant}
\acro{PKI}[PKI]{public key infrastructure}
\acro{PoW}[PoW]{proof of work}
\acro{QWAC}[QWAC]{qualified website authentication certificate}
\acro{SSI}[SSI]{self-sovereign identity}
\acro{URL}[URL]{uniform resource locator}
\acro{VC}[VC]{verifiable credential}
\acro{VP}[VP]{verifiable presentation}
\acro{W3C}[W3C]{world wide web consortium}
\acro{ZKP}[ZKP]{zero-knowledge proof}

\end{acronym}

%% file: 00-Abstract.tex
\begin{abstract}
The digital transformation of the medical sector requires solutions that are convenient and efficient for all stakeholders while protecting patients' sensitive data. One example that has already attracted design-oriented research are medical prescriptions. However, current implementations of electronic prescription management systems typically create centralized data silos, leaving user data vulnerable to cybersecurity incidents and impeding interoperability. Research has also proposed decentralized solutions based on blockchain technology, but privacy-related challenges have often been ignored. We conduct design science research to develop and implement a system for the exchange of electronic prescriptions that builds on two blockchains and a digital wallet app. Our solution combines the bilateral, verifiable, and privacy-focused exchange of information between doctors, patients, and pharmacies through verifiable credentials with a token-based, anonymized double-spending check. Our qualitative and quantitative evaluations as well as a security analysis suggest that this architecture can improve existing approaches to electronic prescription management by offering patients control over their data by design, a high level of security, sufficient performance and scalability, and interoperability with emerging digital identity management solutions for users, businesses, and institutions. We also derive principles on how to design decentralized, privacy-oriented information systems that require both the exchange of sensitive information and double-usage protection. 
\end{abstract}

%% file: 01-Introduction.tex
\section{Introduction}
\label{sec:introduction}

The ongoing digital transformation of the healthcare sector affects its stakeholders in various ways. Healthcare providers, public institutions, and patients alike face a constant pressure to develop and use digital tools in the healthcare sector. As recent developments caused by the Covid-19 pandemic indicate, this pressure is often exacerbated by the need to balance privacy requirements and adequate digital healthcare provisioning~\cite{guggenberger2021covid}. However, this apparent dichotomy appeared before the pandemic, too, for example in the context of digital health records or electronic medical prescriptions~\cite{meena2019preserving, thatcher2018pharmaceutical}. Medical prescriptions refer to authorizations issued by qualified healthcare practitioners that allow patients to obtain medication and services for medical treatment. These authorizations typically manifest as physical, paper-based documents signed or sealed by a qualified physician, which patients present to pharmacies or health service providers. However, such paper-based medical prescriptions suffer from various drawbacks. For example, due to their format, they proved to be slow to process~\cite{seaberg2021use} and susceptible to manipulation, unauthorized reproduction, and errors~\cite{mundy2002system}. Moreover, physical prescriptions can hardly integrate with telemedicine, which has increased by a factor of up to~78 during the Covid-19 pandemic and has stabilized at a level that is~38~times higher than pre-Covid~\cite{mckinsey2021telemedicine}. Paper-based prescriptions also impede automatic checks for cross-reactions of pharmaceuticals~\cite{aldughayfiq2020digital} and seamlessly claiming reimbursement from health insurances. Consequently, several attempts to introduce medical prescriptions in an electronic format have emerged. These digital references or documents allow for automatic validity checks and are typically stored in databases run by parties that are regarded as trustworthy to prevent fraud and the abuse of sensitive data~\cite{aldughayfiq2020digital}. Thus, existing approaches to \acp{e-prescription} rely on highly centralized infrastructures and data silos, such as current efforts in Germany illustrate~\cite{McKinseyeRezept}. While this design is relatively simple, avoids the double-usage of e-prescriptions, and addresses confidentiality requirements through access control enforced by the trusted third party, the corresponding data silos are attractive targets for attackers aiming to capture sensitive health information on a large scale~\cite{le2016state,breaches2020}. Moreover, they pose the socio-economic threat of creating monopolies or oligopolies~\cite{wu2018toward} and ethical issues associated with privacy concerns~\cite{aldughayfiq2020digital}. Also, centralized implementations of \acp{e-prescription} are often not interoperable as they create lock-in effects. In consequence, patients, healthcare practitioners, or pharmacies need to purchase proprietary and expensive software~\cite{bruthans2020state}.

To eliminate these drawbacks of centralized approaches, researchers recently proposed alternative solutions based on decentralized infrastructures~\cite{Siqueira2021,sonkamble2021}. These approaches rely on distributed data storage and aim to provide higher security, privacy, and interoperability than centralized approaches~\cite{stafford2020health}. As such, decentralized solutions address several issues of both paper-based and centralized electronic approaches to medical prescriptions. Extant literature often uses a blockchain as the underlying decentralized infrastructure. Its function as a synchronized single source of truth allows for the avoidance of, e.g., manipulations or the double-spending of \acp{e-prescription}~\cite{zhang2019security}.
Nonetheless, several issues remain, and new problems emerge in blockchain-based approaches. In particular, privacy concerns are exacerbated due to replicated data storage as well as the immutability of blockchains that renders deletion practically impossible~\cite{he2019blockmeds,stafford2020health}. This affects not only patients' confidentiality requirements but also compliance with associated regulatory requirements, such as the EU's \ac{GDPR}, the \ac{CCPA} or the US~\ac{HIPAA}. Furthermore, interoperability often remains a problem, as the suggested implementations are not built on common standards, and sensitive data is typically exchanged using third-party infrastructures. Besides, the existing suggestions often impose the management of cryptographic keys directly on the users or rely on yet another smartphone app (or even device) that users need to carry and use specifically for \acp{e-prescription}.

In the last years, an alternative approach for exchanging verifiable data in a decentralized way has emerged through portable digital identities managed in so-called ``digital wallets''~\cite{sedlmeir2021digital,weigl2021ssi}. This paradigm is often termed \ac{SSI}~\cite{ehrlich2021self,sedlmeir2021digital}. \ac{SSI} offers standards and protocols for end-to-end encrypted bilateral communication and practices for selective and verifiable information disclosure based on digital certificates~\cite{ferdous2019search,W3C2019VCredentials}. However, preventing the double-spending of \acp{e-prescription} that are purely based on digital certificates is not possible solely through bilateral communication, as one pharmacy cannot know whether an \ac{e-prescription} has already been presented and redeemed at another pharmacy. We hence propose that the combination of blockchain technology for double-spending prevention and \ac{SSI} digital wallets for the verifiable exchange of sensitive \ac{e-prescription} data and key management potentially offers properties solving current solutions' shortcomings regarding interoperability, performance, scalability, and security to \ac{e-prescription} management. Summarizing, paper-based prescriptions are susceptible to forgery, inefficiency, and do not integrate with digital services. Digitizing prescriptions through centralized technical infrastructures poses socio-economic, interoperability, and security issues. Decentralizing \acp{e-prescription} through blockchain technology has proven to be susceptible to privacy-related issues and did not specify a suitable user interface. Digital wallets based on the principles of \ac{SSI} have been suggested in fields with similar requirements, such as event ticketing~\citep{feulner2022ticketing}, and they may
be suitable to also solve this issue~\cite{sartor2022ux}. Thus, we pose the following research question:
\textit{How to design and implement a decentralized system for \ac{e-prescription} management using blockchain technology and digital wallets?} To answer this research question, we follow the design science research paradigm. Our evaluation finds that our solution overcomes the shortcomings of existing approaches regarding interoperability, performance, scalability, and provides a sufficient level of security.

%% file: 02-Background.tex
\section{Theoretical Background}
\label{sec:background}

\subsection{E-Prescriptions}
\label{sec:EPresc}
In the UK alone, nearly 1.5 million medical prescriptions are processed on a daily basis~\cite{aldughayfiq2020digital}. Similarly, 11.6 prescriptions were filled in US-pharmacies per capita in the year 2019~\cite{Kaiser2021Prescriptions}. The value of prescription pharmaceuticals prescribed per year globally accounts for approximately \$1~trillion~\cite{Mattke:2019:BlockchainPharma}. To account for limitations of existing approaches to handling the increasing number and value of medical prescriptions, researchers and practitioners alike aim to optimize the process of prescribing and dispensing pharmaceuticals. Accordingly, \ac{e-prescription} systems have been introduced, which can help to increase both the accuracy and the efficiency of provisioning medication to ensure patients’ safety while streamlining processes~\cite{aldughayfiq2020digital, grossman2012transmitting, thatcher2020rxblock}. Submitting and exchanging prescription information digitally is considered to reduce medication errors, minimize overhead due to paperwork and, thus, time resources needed for administrative tasks, additional to increasing patients’ convenience~\cite{aldughayfiq2020digital, grossman2012transmitting}. Accordingly, \ac{e-prescription} management systems enable the creation of prescriptions, their transmission to a pharmacy, and their validation and usage at a pharmacy in a solely digital manner.

However, \ac{e-prescription} systems come with several challenges. While paper-based prescriptions can be physically invalidated upon redemption, \acp{e-prescription} must have a sophisticated mechanism for preventing double-spending~\cite{kierkegaard2013prescription}, as copying electronic documents has close to zero marginal costs. This requires some degree of transparency about whether an \ac{e-prescription} has been already redeemed, particularly when the ecosystems involves several different stakeholders such as many independent pharmacies and doctors among which the patient can choose. These transparency requirements, however, are to be balanced with the protection of sensitive information, for example, patients' data according to the \ac{GDPR} within the~EU. This includes aspects like the right to erasure (``Right to be forgotten''), meaning that users can request their data to be deleted at any point in time (Article~17~\ac{GDPR})~\cite{GDPR2016}. Besides, the data specified in an \ac{e-prescription} must also be secure against manipulation by unauthorized parties. A further challenge for \ac{e-prescription} systems is posed by the need for standardization, as data processing and exchange between many stakeholders requires interoperable systems~\cite{kierkegaard2013prescription}.

Countries such as the USA, Australia, the UK, Spain, and Denmark have already introduced \ac{e-prescription} systems at scale~\cite{aldughayfiq2020digital, bruthans2020state}. The respective systems are usually not standalone products but embedded in more comprehensive e-health systems including, e.\,g., patients' electronic health records. A comparative study conducted by~\citet{aldughayfiq2020digital} found that most implementations build on designs employing a central database for storing \acp{e-prescription}. This design choice can effectively enforce access control and compliance with business logic (such as double-spending prevention) and provide fine-grained access management rules to address confidentiality requirements. However, there are also considerable risks. First, a centralized approach represents a single point of failure, meaning that the system is more vulnerable to attacks~\cite{wu2018toward}. This is also illustrated by a considerable number of privacy breaches in centralized systems in the past~\cite{Papageorgiou}. 
Second, patients cannot be sure that the operator of a centralized system or database does not sell sensitive data to third parties. Sensitive data is available for any participant integrated in the infrastructure with the necessary access permissions~\cite{aldughayfiq2020digital}, and the patient needs to rely on the effectiveness of regulation and audits to enforce data protection. Third, centralized solutions challenge the interoperability of systems. For example, \citet{bruthans2020state} found that the authentication mechanisms used for \ac{e-prescription} systems in Europe are often incompatible. Fourth, network effects carry the risk of centralizing control over data and operations and, thus, can cause monopolies, which hinders productive collaboration in the long run~\cite{hein2019digital}. 

Decentralized systems aim to address some of these problems. For example, Surescripts, an \ac{e-prescription} management system in the~USA, allows entities such as caregivers to access patient information from patients' pharmacies and health insurance companies through the system~\cite{aldughayfiq2020digital}. However, such approaches impose significant responsibility regarding IT security and access management on pharmacies and health insurances, which may be particularly problematic for small organizations. Furthermore, in Surescripts, patients have to choose a specific dispensing pharmacy when their \ac{e-prescription} is created, which decreases users' flexibility and convenience. Consequently, researchers have continued searching for better solutions, and in recent years, they have considered blockchain technology~\cite{wu2018toward}.

\subsection{Blockchain}
\label{subsec:blockchain}

Since the emergence of Bitcoin~\cite{Nakamoto:2008:Bitcoin}, an increasing number of researchers, companies, and government agencies have pointed out the technology's potential beyond financial applications to provide decentralized digital infrastructures and improve cross-organizational processes~\cite{hughes2019blockchain,fridgen2018cross}. This has also led to the emergence of a large number of architectural proposals and implementations at varying stages of maturity, especially in the healthcare sector~\cite{mcghin2019blockchain}.
At its core, blockchain technology provides a distributed and replicated append-only database that groups transactions in blocks on each node of a peer-to-peer network~\cite{butijn2020blockchains}. Aiming to obliterate the need for a central trusted authority~\cite{Nakamoto:2008:Bitcoin}, the nodes of the peer-to-peer network repeatedly agree on the state of the system by following a consensus protocol~\cite{chanson2019blockchain, glaser2017blockchain}. Each block in a blockchain is linked to the previous block through a cryptographic hash pointer. The blocks, therefore, form a chain, creating a tamper-resistant historical data record~\cite{butijn2020blockchains}. Thus, blockchains create a single point of truth between all participants in the distributed ledger~\cite{rossi2019blockchain}. As such, blockchain technology can create neutral platforms~\cite{amend2021we, hoess2021blockchain,sedlmeir2022transparency, voelter2021trust} that naturally improve the interoperability between individual organizational solutions. Previous research has already analyzed the benefits of blockchain for interoperability in the health sector~\cite{dagher2018ancile,gordon2018blockchain}.
Furthermore they provide a solution to the double-spending problem in decentralized systems~\cite{Nakamoto:2008:Bitcoin}. Due to their technical structure, blockchain systems provide several important properties. Resulting from the resistance against crashes or the malicious behavior of a small number of nodes, blockchain systems are highly available and decentralized digital infrastructures~\cite{amend2021evolution}. To participate in consensus or to interact with the peer-to-peer network and authorize transactions, users of a blockchain system must authenticate using public-key cryptography. As a result, blockchain systems also offer an integrated public key infrastructure. 

To account for the requirements of different use cases, various concepts of blockchain technology have emerged. A prominent conceptualization by~\citet{peterspanayi2016blockchain} distinguishes blockchains along two dimensions. First, transactions in public blockchains are publicly visible, while transactions in private blockchains are only visible to the parties that run the network's nodes. Second, permissionless blockchain systems allow anyone to participate in consensus by proposing new blocks, while this right is retained to authorized parties only in permissioned blockchain systems. As the technology has gained increasing prominence, developers introduced smart contracts, which can be defined as scripts that are executed redundantly on the nodes' virtual machines~\cite{buterin2014ethereum,lockl2020toward}. Smart contracts enable a large variety of transactions that go well beyond the transfer of cryptocurrencies~\cite{beck2018blockchain}. One specific application area of smart contracts is the exchange of a broad variety of digital assets. These so-called \emph{tokens} are value containers that represent digital or non-digital, scarce objects and that can be transferred among the participants in a blockchain system~\cite{oliveira2018token, pilkington2016blockchain}. The opportunities related to the ``tokenization'' of physical and digital objects are considered an essential trend for the economy~\cite{sunyaev2021token}. Tokens' inherent value can also help coordinate processes among mutually distrusting parties or in fluid organizations\cite{lee2020data,schirrmacher2021token}. 

Nonetheless, several tradeoffs and challenges remain when using blockchain systems~\cite{kannengiesser2020trade}. While energy consumption is only problematic for a specific consensus mechanism, namely proof of work~\citep{sedlmeir2020energy}, blockchains in general exhibit challenges regarding scalability and data visibility due to the inherent replicated storage and execution of transactions~\cite{kolb2020core,kannengiesser2020trade}. Specifically excessive information exposure implies significant challenges both from the perspective of natural persons' data protection as well as organizations' sensitive business data. This issue is aggravated by blockchains' immutability guarantees that inhibit the retrospective deletion of sensitive information accidentally stored on a blockchain~\citep{schellinger2022gdpr}. Blockchains' inherent pseudonymization through addresses representing public keys does not considerably mitigate this problem, as the aggregation of information from different domains can provide sufficient means for de-anonymization~\cite{reid2013analysis}. While permissioned blockchains partially address this problem through restricting access to selected organizations running the network's nodes, still all of these participants can see each other's transactions by design. Thus, from the perspective of an attacker that intends to retrieve sensitive information, a permissioned blockchain is a centralized system with yet more attack vectors~\cite{platt2021}. ``Private transactions'' that contain only hashed or encrypted information have been implemented in permissioned blockchains to address enterprises' requirements (e.\,g., in Hyperledger Fabric and Quorum); however, this approach also restricts the functionality of smart contracts, as these typically cannot run on obfuscated data. Trusted execution environments and advanced cryptographic tools such as \acp{ZKP}, multiparty computation, and (fully) homomorphic encryption can be used to bring some of this functionality back, but can come at significant complexity and often have their own performance issues~\cite{zhang2019security, sedlmeir2021labeling}. Consequently, the use of blockchains should be well-considered when it comes to the processing of sensitive information.

\subsection{Self-Sovereign Identity and Digital Wallets}
\label{subsec:SSI}

\Ac{SSI} aims to facilitate verifiable digital identities for organizations, end users, and networked machines that are not tied to a certain place or organization and that can be used across domains with the identity owners' consent~\cite{allen2016pathtossi}. \ac{SSI} involves three distinct types of entities~\cite{muhle2018survey}: the issuer of an identity document, the holder of the respective document, and the verifier of properties described in the document. An analogy from the physical world serves as an illustration of the basic interactions~\cite{sedlmeir2021digital}: An \ac{SSI} system builds upon digital representations of tamper-resistant physical documents like ID~cards or driver's licenses~\cite{DBLP:journals/csm/AvellanedaBBBDD19}. Appropriate organizations, such as government authorities, issue the respective documents to their holders, who subsequently store them in a physical infrastructure of their choice, such as a wallet~\cite{weigl2021ssi}. Such documents typically follow specific schemas, such as watermarks and attribute names, which have been made public by their issuer. As a result, the integrity of such documents can be verified by third parties in bilateral interactions with their holders. In this system, the issuer's trustworthiness is important from the verifier's perspective.

The building blocks and principles of an \ac{SSI} system can be derived from this analogy. Tamper-resistant physical documents relate to \acp{VC}, which are cryptographically signed digital objects containing claims about their holder's identity and authorizations~\cite{ehrlich2021self, Preukschat2019SSI, W3C2019VCredentials}. Holders store these \acp{VC} in an (for end users typically mobile) application called ``digital wallet''~\cite{sartor2022ux}. To prove properties, or claims, described in their \acp{VC} to a verifier, holders generate \acp{VP}. These \acp{VP} are tamper-proof attestations derived from one or multiple \acp{VC} to address the requirements posed by a verifier~\cite{Hardman2019verifpresentation, Preukschat2019SSI, W3C2019VCredentials}. This can either be achieved by presenting the signed credentials themselves, e.\,g., a JSON~Web Token, or by creating a cryptographic \ac{ZKP}, for example from an anonymous credential~\cite{camenisch2001credentials, hardman2020savvy}. The latter method allows data minimization in the form of selective disclosure, where only a subset of the attributes contained in a \ac{VC} can be revealed. 

Entities within an \ac{SSI} system often use so-called \acp{DID} as identifiers, which are unique and follow a standardized schema~\cite{W3C2020DID}. They also serve to create a standardized end-to-end encryption between two parties. \acp{DID} corresponding to public institutions can be made publicly discoverable as an alternative to \ac{CA}-based binding of public keys, domain names, and IP~addresses. To avoid unnecessary causes of correlation, it is recommendable for natural persons to use a new, private (``peer'') \ac{DID} in each interaction~\citep{schlatt2021designing}. While these building blocks provide a solid foundation for an \ac{SSI} system, a neutral infrastructure is necessary: information about issuers of \acp{VC}, such as their current signing keys, and revocation-related information must be publicly available to verify the correctness of \acp{VP}. By proving knowledge of the issuer's digital signature and non-inclusion of their \ac{VC} in a public but privacy-protecting revocation registry (in the form of a cryptographic accumulator), holders can convince a verifier that their \ac{VC} has not been revoked without having to contact the credential issuer~\cite{schlatt2021designing}. Furthermore, schemas of \acp{VC} must be publicly available to verify the integrity of \acp{VP}. Due to its properties as a decentralized and highly available data structure, blockchain technology is often used for this purpose~\cite{ferdous2019search,muhle2018survey}. Lately, the concept of \ac{SSI} and digital wallets has been applied in a variety of endeavours~\cite{kuperberg2020blockchain,schlatt2021designing,feulner2022ticketing}. For example, the current efforts of the European Union to improve the legislation as well as the realization of \ac{eIDAS} feature many parallels to \acp{SSI}-based systems~\cite{eidas2022eval} . One of the proposed use cases to be developed in the context of \ac{eIDAS}~2.0 specifically represents \acp{e-prescription}~\cite{EU2022eIDASArchitecture}.

%% file: 03-Method.tex
\section{Method}
\label{sec:method}

\subsection{Design Science Research Approach}
\label{subsec:dsr}

We answer our research question by employing a \ac{DSR} approach. The goal of \ac{DSR} is to design, develop, and evaluate IT artifacts as solutions to practical challenges, thus aiming to solve real-world problems~\citep{march1995design, hevner2004design}. \ac{DSR} faces the dichotomy of proposing a feasible solution to practice, whilst providing a valid contribution to theory at the same time~\citep{baskerville2018design}. We aim to solve this issue by designing, implementing, and evaluating a system for \acp{e-prescription} based on blockchain technology and digital wallets. The resulting IT artifacts build on foundations from previous research. To infer a contribution to theory, we subsequently derive \acp{DP} to offer more generalizable knowledge on the implications of designing and developing IT artifacts with similar requirements~\citep{gregor2013positioning} such as privacy protection and double-spending prevention.
We adhere to the \ac{DSR} model proposed by~\citet{peffers2007design} to guide our research, which consists of six partly overlapping as well as iteratively conducted steps. The initial step starts the process by defining a research problem of practical relevance. As laid out in section~\ref{sec:background}, current solutions for \acp{e-prescription} are prone to security incidents, lack interoperability, provide socio-economic risks, and offer opportunities for fraud. Thus, the design of systems for \acp{e-prescription} solving these issues is a problem of practical relevance. In the second step, we derive requirements for our solution by reviewing existing proposals for implementing \acp{e-prescription} identified in a structured literature review. We define~8~design objectives for our proposed solution, which we present in section~\ref{subsec:design_objectives}. Adhering to these design objectives, we develop a system architecture and instantiate it by implementing a prototype based on blockchain technology and digital wallets\footnote{Our implementation is available at \url{https://github.com/JSedlmeir92/e-Prescriptions}.}. We thus contribute both an architectural design as well as a prototype as IT~artifacts. To evaluate the fulfilment of our design objectives and to highlight advantages as well as remaining shortcomings, we evaluate the developed IT~artifacts quantitatively as well as qualitatively along the criteria defined by the design objectives~\citep{sonnenberg2012evaluations}. In a sixth step, we present our artifact in this publication.

\subsection{Literature Review}
\label{subsec:litrev}

\begin{figure}[!t]
    \centering
    \includegraphics[width=\linewidth, trim=4.5cm 5.5cm 2cm 0cm, clip, page=1]{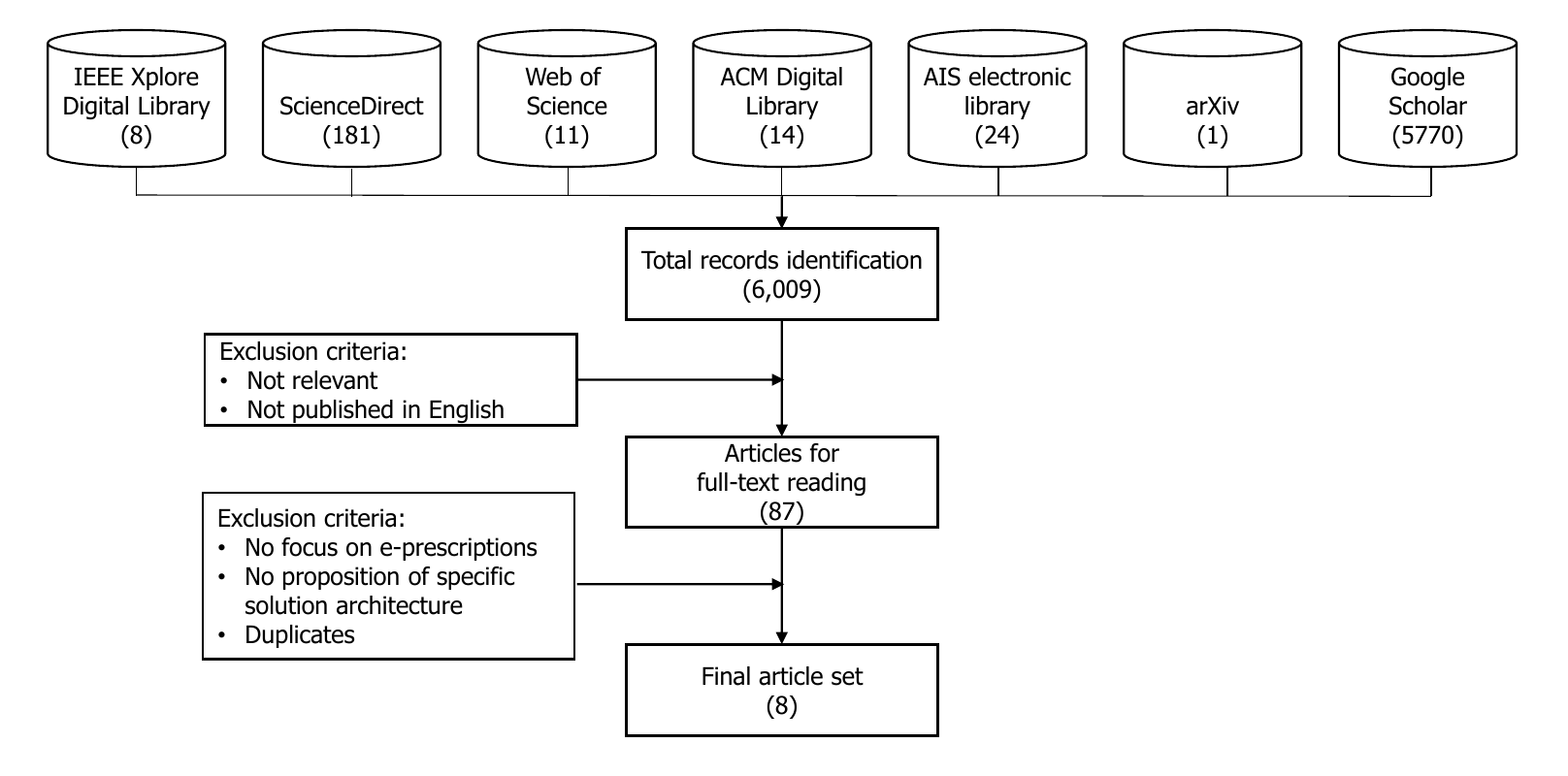}
    \caption{Systematic literature review for the search string ``blockchain AND prescription AND health''.}
    \label{fig:literature_review}
\end{figure}

We conducted a structured literature review to identify relevant work investigating the potential of decentralized approaches for \acp{e-prescription} management systems following the guidelines proposed by~\citet{Kitchenham07guidelinesfor}. Accordingly, we derived our search string on the basis of our research question~\cite{Kitchenham07guidelinesfor}: ``blockchain AND prescription AND health''. We decided against using synonyms for blockchain technology as it is regarded as the most prominent concept associated with decentralization and often serves as a solution for implementing \acp{e-prescription} in a decentralized way~\cite{he2019blockmeds, tanwar2020blockchain, thatcher2020rxblock}. We screened the databases ACM~Digital~Library, AISeL, arXiv, IEEE~Xplore, Google~Scholar, ScienceDirect and Web~of~Science as they represent the prevailing databases in the computer science and information systems research domain. The initial search yielded~6,009 results in total (see figure~\ref{fig:literature_review}).

We included research in English language only. During title screening, we excluded papers that did not mention blockchain and health (or synonyms) in their titles to ensure the relevance of our article set. As Google Scholar yielded~5770 results, we screened the titles of all results until no papers related to our research question could be identified for 50~results in a row, sorted by relevance. Subsequently, we screened the contents of the remaining papers. First, we excluded papers that did not focus on solutions for \acp{e-prescription}. Second, following the goal of our literature review to identify existing propositions for \acp{e-prescription}, we also excluded papers that do not cover a specific solution architecture or that give guidance on architecture design. Last, we excluded duplicates from our article set. We found that although blockchain is often mentioned as a viable technology for the health care sector in general~\cite{engelhardt2017healthcarebc,katuwal2018applications, seitz2020opportunities}, the literature dismissed challenges related to the privacy of medical data~\cite{Siqueira2021}, mainly neglecting privacy-by-design in blockchain-based health systems~\cite{sonkamble2021}. Moreover, extant literature only rarely explicitly proposes architectures for the management of \acp{e-prescription}. In contrast, the current literature body places its focus on blockchain-enabled medical supply chains~\cite{jamil2019novel, Mattke:2019:BlockchainPharma,ruby2020authentic}, electronic health records~\cite{chenthara2020healthchain,zhang2018blockchain}, and the management of medication histories~\cite{kim2019patient, raghavendra2019can}. Furthermore, while some authors address \acp{e-prescription} explicitly, they focus on implementations and requirements in a specific country, hindering the generalization of their findings~\cite{mahatpureelectronic}.

%% file: 04-Requirements.tex
\section{Related Work and Design Objectives}
\label{sec:related_work}

\subsection{Related Work}
\label{subsec:related_work}

Through our literature review, we identified seven papers in total that explicitly propose or survey blockchain-based solutions for \ac{e-prescription} processes. After we had identified the final article set, we extracted information about the requirements mentioned by the authors, the proposed solution architecture, as well as related advantages and challenges. To ensure intercoder reliability, every paper was coded by at least two authors. In the following, we will analyze these papers depicted in table~\ref{tab: Results1} in detail.

  \renewcommand{\arraystretch}{1.5}
  \begin{table}[!b]
    \centering
    \resizebox{0.95\linewidth}{!}{
    \vspace{0cm}
    \setlength{\aboverulesep}{0pt}
    \setlength{\belowrulesep}{0pt}
    \begin{tabular}{p{3cm}|c|p{12cm}|c}
        \toprule
        Authors & Year & Title & Ref. \\
        \midrule
        Cadoret et al. & 2020 & Proposed Implementation of Blockchain in British Columbia's Health Care DataManagement &~\cite{cadoret2020proposed}\\
        Gropper & 2016 & Powering the Physician-Patient Relationship with HIE of One Blockchain Health IT &~\cite{gropper2016powering}\\
        He et al. & 2019 & BlockMeds: A Blockchain-Based Online Prescription System with Privacy Protection &~\cite{he2019blockmeds}\\
        Li et al. & 2019 & DMMS:  A  Decentralized  Blockchain  Ledger  for  the  Management of Medication Histories & \cite{li2019dmms}\\
        Meena et al. & 2019 & Preserving Patient's Privacy using Proxy Re-Encryption in Permissioned Blockchain & \cite{meena2019preserving}\\
        Thatcher and Acharya & 2018 & Pharmaceutical Uses of Blockchain Technology & \cite{thatcher2018pharmaceutical}\\
        Thatcher and Acharya & 2020 & Towards the Design of a Distributed Immutable Electronic Prescription System & \cite{thatcher2020rxblock} \\
        Ying et al. & 2019 & A Secure Blockchain-based Prescription Drug Supply in Health-Care Systems & \cite{ying2019secure} \\
        \bottomrule
    \end{tabular}
    } 
    \vspace{0cm}
    \caption{Results of the structured literature review (conducted in January 2021).}
    \label{tab: Results1}
    \end{table}

Many publications exploring the advantages of blockchain technology for \acp{e-prescription} do not address the privacy-related challenges of blockchain technology in their suggested solutions. For example,~\citet{thatcher2018pharmaceutical} as well as~\cite{thatcher2020rxblock} propose an \ac{e-prescription} management system that integrates a blockchain-based ``immutable database of prescriptions''. While the proposed design ensures tamper-resistance and accountability for processing \acp{e-prescription}, the authors acknowledge that it dismisses patients' privacy~\cite{thatcher2018pharmaceutical}, thereby highlighting an essential requirement for \acp{e-prescription}. \citet{he2019blockmeds} also acknowledge privacy-related issues in current approaches to decentralized \ac{e-prescription} systems and suggest a permissioned blockchain for the management of \acp{e-prescription}. However, their proposed architecture still stores sensitive patient data on-chain and hence leaves them more vulnerable to cyber-attacks than they would be in a centralized database, as we discussed in section~\ref{subsec:blockchain}. Similarly, \citet{cadoret2020proposed} suggest storing prescription data in private, permissioned blockchains. They also propose using \acp{ZKP} for addressing privacy challenges if a public blockchain was used, but do not provide details how the associated access management would work or on which layer the sensitive data could be exchanged. In both solutions, patients do not have control over which data is shared. This problem is addressed by~\citet{li2019dmms} and \citet{meena2019preserving}, who suggest systems which allows patients to decrypt data that is saved on the ledger upon request by pharmacies. As a result, stakeholders other than the patient are unable to make sense of the encrypted data. Yet, as the decrypted data is shared bilaterally between the patient and the pharmacy, the purpose of blockchain is limited to facilitating data integrity checks. Simpler means to achieve data integrity, such as digital signatures, are not considered. Moreover, the permanent storage of encrypted data on a blockchain may become an attack vector in the future due to the increase of computing power over time and the potential availability of sufficiently powerful quantum computers within the next decades~\citep{schellinger2022gdpr}. This concern is manifested by ongoing controversial debates in the~EU~whether or not encrypted data should be classified as personally identifiable~\cite{finck2018blockchains}. Both publications do not suggest how users' cryptographic keys get managed and how patients are enabled to interact with various involved stakeholders. Finally,~\citet{ying2019secure} propose an architecture that ensures privacy throughout the process of dispensing drugs, but they omit the integration of central stakeholders such as the users in their considerations and, thus, cannot give users control over which data is being shared between doctors and pharmacies.

Our structured literature review includes two publications whose concepts build upon \ac{SSI} in the digital healthcare context. While \citet{cadoret2020proposed} briefly outline \ac{SSI} as an authentication mechanism for users of blockchain-based medical services, \citet{gropper2016powering} propose a more extensive \ac{SSI}-based system for overcoming the current challenges of electronic medical records. However, the latter authors do not provide details on how multiple usages of \acp{e-prescription} can be prevented. If there is only one pharmacy where the \ac{e-prescription} can be spent, a serial number in the \ac{VC} that needs to be revealed can be used to prevent repeated redemptions. In the case of multiple pharmacies, one would either need to synchronize the serial numbers of redeemed \ac{VC} among all pharmacies or require interactions between pharmacies and doctors, where the pharmacy redeeming the prescription asks the doctor to revoke the credential. This approach would, however, introduce significant additional complexity.

\subsection{Design Objectives}
\label{subsec:design_objectives}

From the structured literature review, we derive a set of design objectives for a decentralized \ac{e-prescription} management system. Both existing paper-based and \ac{e-prescription} processes demonstrate certain requirements that must be fulfilled. For example, the current redemption process of paper-based prescriptions ensures that each prescription may be redeemed only once. Furthermore, the strengths and weaknesses of decentralized approaches proposed in related work allowed us to derive additional requirements. 

Table~\ref{tab: Requirements} summarizes the design objectives that we found in our literature review. 
\renewcommand{\arraystretch}{1.5}
\begin{table}[!tb]
\centering
\resizebox{\linewidth}{!}{
\vspace{0.25cm}
\begin{tabular}{r|p{2.25cm}|p{12cm}|p{2.75cm}}
    \toprule
    Nr. & Requirement & Context & References \\
    \midrule
    R.1 & Disclosure \mbox{control} & As personal health data is highly sensitive, users' privacy must be ensured. Users can choose with whom to share what parts of their \ac{e-prescription}. &  \cite{cadoret2020proposed}, \cite{gropper2016powering}, \cite{he2019blockmeds}, \cite{li2019dmms}, \cite{meena2019preserving}, \cite{thatcher2018pharmaceutical}, \cite{thatcher2020rxblock} , \cite{ying2019secure}\\
    R.2 & Decentralization & To prevent the abuse of patient data through operators of centralized solutions and large-scale data breaches, no centralized data silos containing patient data must exist. A decentralized solution can also ensure high availability guarantees, which is important for the functionality of a critical system. & \cite{cadoret2020proposed}, \cite{gropper2016powering}, \cite{he2019blockmeds}, \cite{li2019dmms}, \cite{meena2019preserving}, \cite{thatcher2018pharmaceutical}\\
    R.3 & Pharmacy \mbox{independence} & To ensure competition and a free choice of pharmacy for patients, the user must not be bound to a specific pharmacy for redeeming the prescription. &\cite{gropper2016powering}, \cite{he2019blockmeds}\\
    R.4 & Key management & The developed system must be convenient to use and should not create unnecessary overhead owing to cryptographic key management for users. & \cite{li2019dmms}, \cite{meena2019preserving}, \cite{ thatcher2018pharmaceutical}\\
    R.5 & Interoperability & Current implementations for \acp{e-prescription} differ significantly, hindering interoperability and thereby adoption as a result. Interoperability could furthermore extend the application areas of \ac{e-prescription} systems. &  \cite{cadoret2020proposed}, \cite{gropper2016powering}, \cite{li2019dmms}, \cite{thatcher2020rxblock} \\
    R.6 & \mbox{Scalability and} performance & The \ac{e-prescription} management system must be able to quickly handle a sufficient number of prescriptions redeemed on a large scale, even at peak times. & \cite{cadoret2020proposed}, \cite{ gropper2016powering}, \cite{he2019blockmeds}\\
    R.7 & Double-spending prevention & To prevent fraud, prescriptions must not be redeemed twice, or not more frequently than the number of times intended in the case of periodic prescriptions. & \cite{thatcher2020rxblock}\\
    R.8 & Verifiability & Pharmacies need to fully automatically verify the integrity and authenticity of \acp{e-prescription} to ensure the correctness of prescription information, the doctor's authenticity as well as the patients' eligibility to receive medicals. & \cite{cadoret2020proposed}, \cite{gropper2016powering}, \cite{he2019blockmeds} \cite{li2019dmms}, \cite{meena2019preserving}, \cite{thatcher2018pharmaceutical}, \cite{thatcher2020rxblock} \\
    \bottomrule
\end{tabular}} 
\vspace{0cm}
\caption{Design objectives for an \ac{e-prescription} management system.}
\label{tab: Requirements}
\end{table}
The derived requirements serve as a basis for the remainder of this paper and inform and guide the design and implementation of our proposed system architecture. They also represent the objectives along which we evaluate our artifact in section~\ref{sec:evaluation}.

%% file: 05-Architecture.tex
\section{Architecture and Implementation}
\label{sec:architecture}

\subsection{Conceptual Architecture}
\label{subec:conceptual_architecture}

\begin{figure}[!tb]
    \centering
    \includegraphics[width=0.9\linewidth, page=2, trim=2cm 0cm 2cm 0cm, clip]{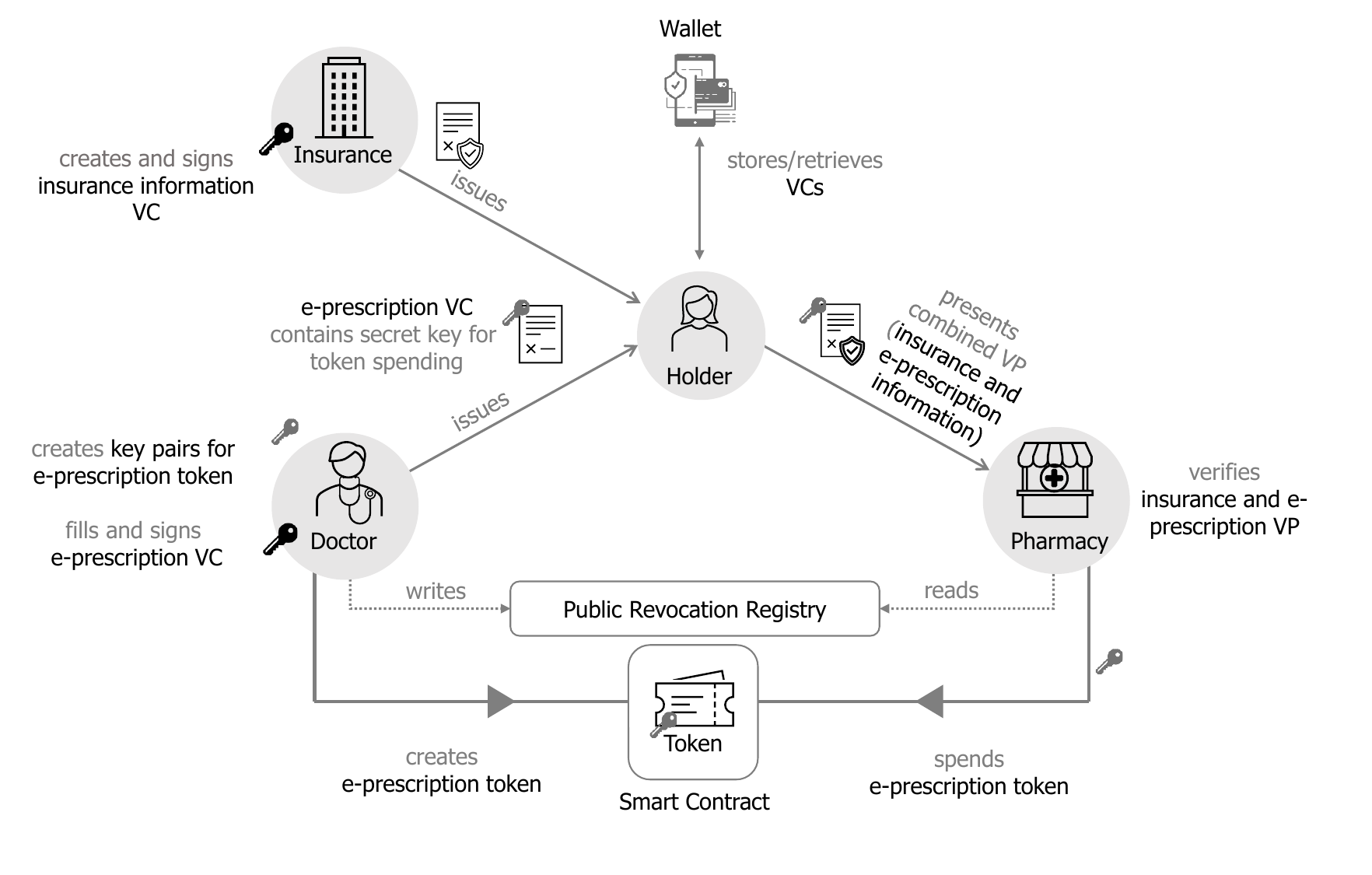}
    \caption{E-prescriptions based on verifiable credentials with anonymized blockchain-based tokens for double-spending checks.}
    \label{fig:stakeholders_and_process}
\end{figure}

Following the \ac{DSR} research paradigm~\citep{hevner2004design,peffers2007design}, we iteratively developed an architecture for the management of \acp{e-prescription} that addresses the previously derived design objectives. We built on the two major paradigms for decentralized digital interactions discussed in section~\ref{sec:background}: On the one hand, we leverage digital wallets for the bilateral and verifiable exchange of sensitive information. On the other hand, we employ blockchain technology to prevent the double-spending of \acp{e-prescription}. Figure~\ref{fig:stakeholders_and_process} features the overall architecture that reflects these two core building blocks.

We split the design process in two cycles: First, we designed the architecture involving the stakeholders \textit{doctor}, \textit{patient}, and \textit{pharmacy} for issuing and redeeming \acp{e-prescription}. Doctors and pharmacies are each running an institutional agent~\citep{schlatt2021designing} to issue \acp{VC} and verify \acp{VP}, respectively. The patient is carrying a digital wallet on their smartphone and interacts bilaterally with the doctors and pharmacies of their choice. No direct communication between doctors and pharmacies is necessary, as the authenticity of \acp{VP} derived from \acp{VC} can be verified through the issuer's signature on the \ac{VC} and revocation registries' accumulator states published on a blockchain or another public data registry. The patient first visits their doctor, who issues an \ac{e-prescription} that includes information such as the patient's and doctor's name, a specification of the prescribed pharmaceutical, and its quantity to the patient's smartphone wallet through a bilateral, \ac{E2E} encrypted communication channel. Subsequently, the patient can present the e-prescription \ac{VC} to a pharmacy, which checks its validity based on the list of doctors and their associated public keys that they trust and retrieves the information to release the right medical.

As we received feedback on the integration of our \ac{e-prescription} solution, we extended our architecture to embed it within the healthcare ecosystem. As a result, patients initially receive a \ac{VC} that attests their identity and health insurance information, such as name, address, and insurance policy number. When connecting with their doctor in a branch or remotely, patients can give a presentation of their health insurance \ac{VC}, which the doctor can use to verify the patients' identity and, thus, to be sure that they issue the \ac{e-prescription} \ac{VC} to the right wallet. Similarly, after having received their \ac{e-prescription}, patients can then connect with their pharmacy, again either remotely or through visiting a branch, and give a combined \ac{VP} of the \ac{e-prescription} \ac{VC} as well as information associated with the health insurance \ac{VC} that they can use to streamline the billing process. To add the respective \ac{e-prescription} to the patient's health record, a receipt could be given to the patient as a \ac{VC}, which they can then present to their insurance maintaining the respective health records. Alternatively, this could be done directly by the pharmacy in a bilateral interaction with information obtained through the \ac{e-prescription} \ac{VC}. We highlight that it is necessary for patients to use a one-time \ac{DID} for communicating with a pharmacy (and a doctor), as otherwise privacy-compromising data analysis could be performed.

To prevent the double-spending of \acp{e-prescription}, we link a blockchain-based token that does not carry sensitive information to the \ac{e-prescription} \ac{VC}. To provide an efficiently and conveniently usable key management associated with the token, the \ac{e-prescription} \ac{VC} includes the private key that can be used to control the token: As the patient trusts the doctor and the pharmacy concerning the prescription that they want to spend, it is not necessary that the patient has exclusive control over the token at any point in time -- the patient can simply carry information that is necessary to retrieve and invalidate the token for double-spend protection \emph{in} the \ac{e-prescription} \ac{VC}. Doctors and pharmacies hence both run a blockchain client, and upon onboarding, a doctor creates a (long-term) key pair (sk$_{\mathrm{doctor}}$, pk$_{\mathrm{doctor}}$) and deploys a smart contract granting the doctor the right to create new tokens. When the doctor wants to issue an \ac{e-prescription}, they generate a new keypair (sk$_{\mathrm{presc}}$, pk$_{\mathrm{presc}}$) that is only used for this one specific prescription. Next, the doctor creates a new token that can only be spent by the controller of pk$_{\mathrm{presc}}$, i.e., by anyone who knows sk$_{\mathrm{presc}}$, in the smart contract by signing an associated transaction with sk$_{\mathrm{doctor}}$. The doctor then includes the smart contract's address and sk$_{\mathrm{presc}}$ in the \ac{e-prescription} \ac{VC}. Consequently, any party that receives a \ac{VP} of an \ac{e-prescription} \ac{VC} that reveals sk$_{\mathrm{presc}}$ can potentially invalidate, i.e., spend, the associated token if they can send it to an entity with write access on the blockchain. Thus, the patient should not reveal this attribute on the \ac{e-prescription} \ac{VC} to a party that they do not trust. Notably, selective disclosure still allows to use the \ac{e-prescription} \ac{VC} for \acp{VP} towards entities the patient does not trust regarding the control of the token.

We employ two separate blockchains for the purpose of storing \ac{SSI}-related information on a specialized infrastructure and implement token-management functions on the other. The rationale behind this design choice is that Hyperledger Indy provides a mature decentralized framework specialized in providing identity management, while Quorum offers wide token management functionalities. Theoretically, the functionalities offered by the Hyperledger Indy blockchain could be implemented on the Quorum blockchain as well. This may even be desirable for managing complexity, albeit initially, several new implementations would be required. Thus, our solution is conceptually reproducible with only one blockchain. Nonetheless, Indy provides a widely accepted standard in other identity ecosystems, such as Germany's IDunion or Canada's Verifiable Organizations Network, which makes integration of the digital wallet in other settings easier. Moreover, as the blockchains manage strictly independent tasks, we consider the complexity due to relying on two blockchain implementations as reasonable for now. Nevertheless, adapting our solution with regards to this aspect could be a long-term improvement goal as soon as the respective tools and frameworks are available within one infrastructure.

Spending the \ac{e-prescription} involves minimal effort, as all necessary information is contained in the \acp{VC}: Upon the verification of a combined \ac{VP} from the \ac{e-prescription} \ac{VC} and health insurance \ac{VC}, the pharmacy can use their blockchain client to invalidate the respective token, as they can control it using sk$_{\mathrm{presc}}$, which they learned from the \ac{VP}. As the transactions in a blockchain are ordered, double-spending is prevented. Even in case the patient tries to spend the \ac{e-prescription} at many pharmacies in parallel, all token-related transactions of pharmacies will be ordered. Thus, only the first transaction will succeed in retrieving the token; all the others will return an error message indicating that the token has already been spent and, thus, the \ac{e-prescription} cannot be redeemed any more. After spending the \ac{e-prescription}, the pharmacy can use the billing information also obtained from the \ac{VP} to enable the automation of their billing processes; which a similar technique that builds on combining a \ac{VC} and a token to avoid that patients can get reimbursement for the same bill multiple times. For all these functionalities, a user does not need a dedicated app, but can use their digital wallet that is familiar from other identification and authentication processes.

\subsection{Implementation}
\label{subsec:implementation}

\begin{figure}[!tb]
    \centering
    \includegraphics[width=0.8\linewidth, page=3]{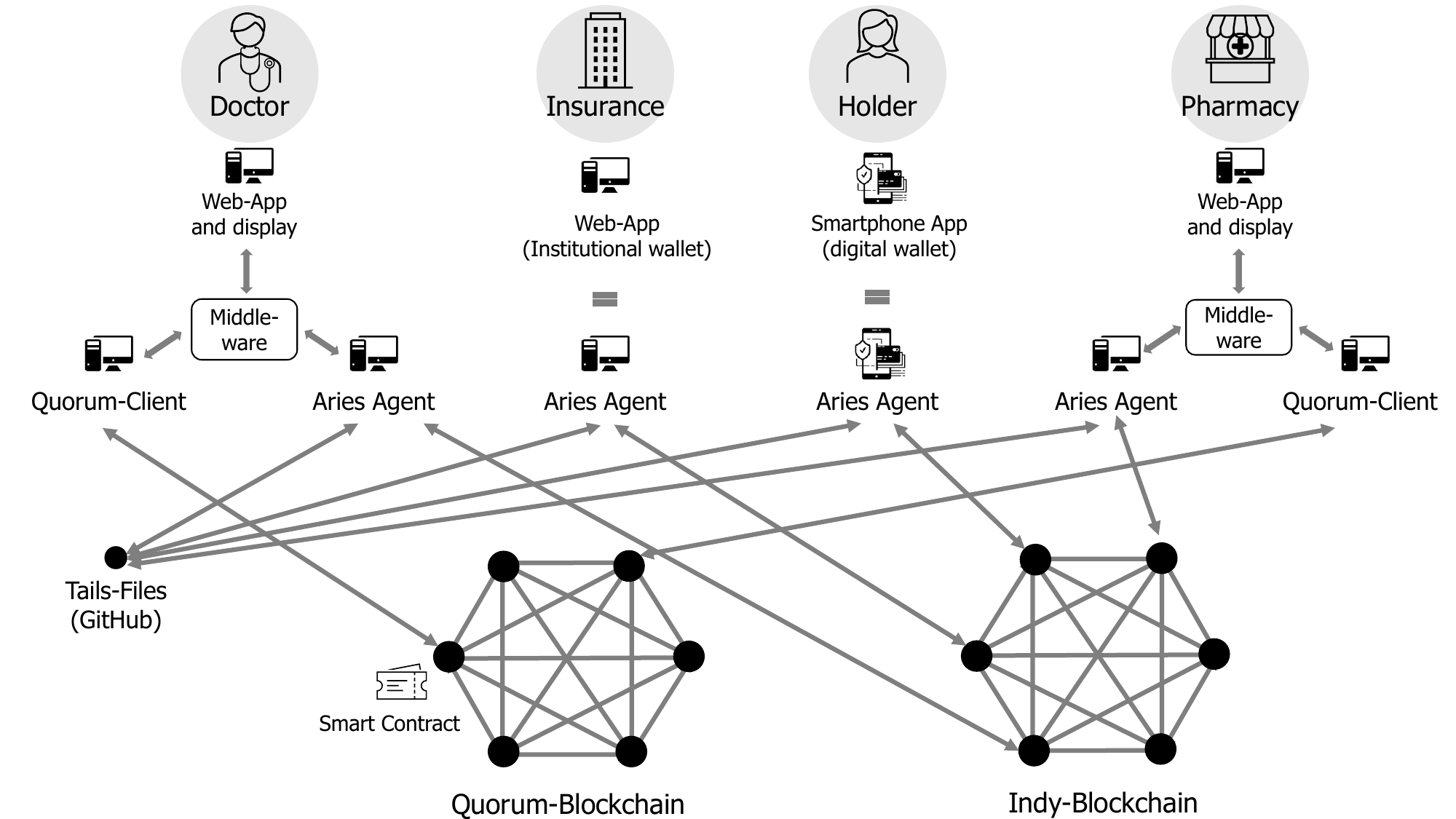}
    \caption{Technical components of the proposed e-prescription management architecture.}
    \label{fig:components}
\end{figure}

We illustrate the components of our implementation of an \ac{e-prescription} system based on our proposed architecture in figure~\ref{fig:components}. We tested this prototype in an isolated environment, without interactions with legacy systems or stakeholders.
Our implementation requires smart contracts to create tokens that prevent the double-spending of \acp{e-prescription}. Consequently, we chose an Ethereum-based blockchain because the Ethereum virtual machine and Solidity represent a sufficiently mature technology stack for smart contract implementation. Furthermore, we aimed for a scalable e-prescription system that provides the necessary level of throughput to operate e-prescriptions for several countries across the European Union while keeping transaction costs negligible. Thus, we decided to use the private Ethereum-based blockchain Quorum. It has been designed for enterprise applications and provides the lowest latency and highest throughput in recent benchmarks~\citep{sedlmeir2021benchmarking,kannengiesser2020trade}. Moreover, due to the use of \ac{IBFT}, it provides a highly secure Byzantine Fault tolerant consensus mechanism, whereas many other private Ethereum implementations only provide crash fault tolerance (e.g., Clique consensus of Geth, Aura consensus of Parity).

We decided for the permissioned option due to its high throughput, low latency, low energy consumption, and low transaction costs. Doctors and pharmacies consequently run Quorum clients for token creation and spending. 
The smart contract that serves to prevent double-spends specifies that a Prescription token must have two attributes, namely its issuer (the doctor's public key) and an indicator of how often it can be spent at most. On instantiation, the contract creates a registry for such prescriptions, i.e., a mapping of public keys (pk$_{\mathrm{presc}}$) to prescription tokens, and specifies that only the creator of the contract (the \emph{admin}, i.e., the doctor) can create new prescription tokens. The smart contract furthermore ensures that only someone who can prove control over a public key associated with a token can spend it, and only as often as initially specified by the doctor in \emph{count}. The smart contract code is provided in figure~\ref{fig:ePrescription_Contract} in the appendix. We also deployed Node.js-based Quorum-clients for doctors and pharmacies, which invoke transactions for creating resp. spending tokens, given the contract address and pk$_{\mathrm{presc}}$ resp. the contract address and sk$_{\mathrm{presc}}$.

We set up a four-node Hyperledger Indy blockchain network that provides the functionality to store a schema defining the attributes of \acp{e-prescription} and health insurance \acp{VC} (``template''), doctor and insurance-specific credential definitions (specifying a doctor's and insurance's signing keys) and revocation registries based on cryptographic accumulators that allow the patient to create \acp{ZKP} of non-revocation in their \ac{VP}.
In addition, we deployed an instance of the \href{https://github.com/hyperledger/aries-cloudagent-python}{Hyperledger Aries} Cloud Agent in Python, which acts as an Indy-client and a RESTful API for bilateral communication,  issuance of \acp{VC} and the verification of \acp{VP}, for doctors, health insurance, and pharmacies. In theory, the proposal could also be implemented with one blockchain alone as long as both the functionalities required for smart contracts as well as identity management are given. However, we chose two different blockchains mainly because they are specifically suitable for the two main tasks that we require: On the one hand, we require token management functionalities, which are currently represented by the Quorum blockchain as outlined above. On the other hand, Hyperledger Indy provides a mature decentralized identity management framework. To the best of our knowledge, there is no alternative implementation as mature as Hyperledger Indy specifically when it comes to compatible digital identity provisioning. Furthermore, we aimed to use an implementation framework for \acp{SSI} that offers users a high level of usability~\cite{sartor2022ux}. Regarding the Aries Cloud Agents, there currently are several different digital wallets are available, which are compatible with the backend (e.g., Trinsic, Lissi, esatus, and ID Wallet).  Additionally, the proposed solution should fulfill a high degree of privacy. The DIDComm protocol used in Hyperledger Indy and Aries relies on~\acp{ZKP} for selective disclosure and set-membership proofs to convince a verifier that an e-prescription is not revoked without revealing a correlatable e-prescription identifier~\cite{schlatt2021designing}.
We also implemented a web-application (frontend) with the \href{https://www.djangoproject.com/}{Django} framework in Python. This web-application can be used for the doctor's onboarding process (deploying the smart contract) as well as connecting to patients, issuing \acp{VC} to them, and requesting a \ac{VP}.
For demonstration purposes of the required functionality for the patient, we employed a smartphone digital wallet app based on the .NET implementation of the Hyperledger Aries protocol. The wallet has been developed by the IT company \href{https://esatus.com/?lang=en}{esatus}~AG and is available for free in the Google Playstore for Android and the App Store for~iOS. It is not built for a specific use case like \acp{e-prescription} but for the generic exchange of verifiable information (such as ID cards, credit cards or diploma) based on standards that are related to the W3C \ac{DID} and \ac{VC} standards~\cite{W3C2019VCredentials, W3C2020DID}. 
Beyond the basic capabilities such as \ac{ZKP}-based selective disclosure and proofs of non-revocation, it additionally supports custom Hyperledger Indy networks, provides backup functionalities, and recently, it was demonstrated that the content of the esatus wallet (including cryptographic keys and \acp{VC}) can be exported to and imported from another wallet implemented by Trinsic~\cite{sovrin2020interop}. Similar wallets are also offered by \href{https://www.evernym.com/}{Evernym} (Connect.me wallet) and the German \href{https://idunion.org/}{IDunion} consortium (Lissi wallet).

\begin{figure}[!tb]
    \centering
    \includegraphics[width=0.9\linewidth, trim=2cm 1cm 2cm 0cm, clip, page=4]{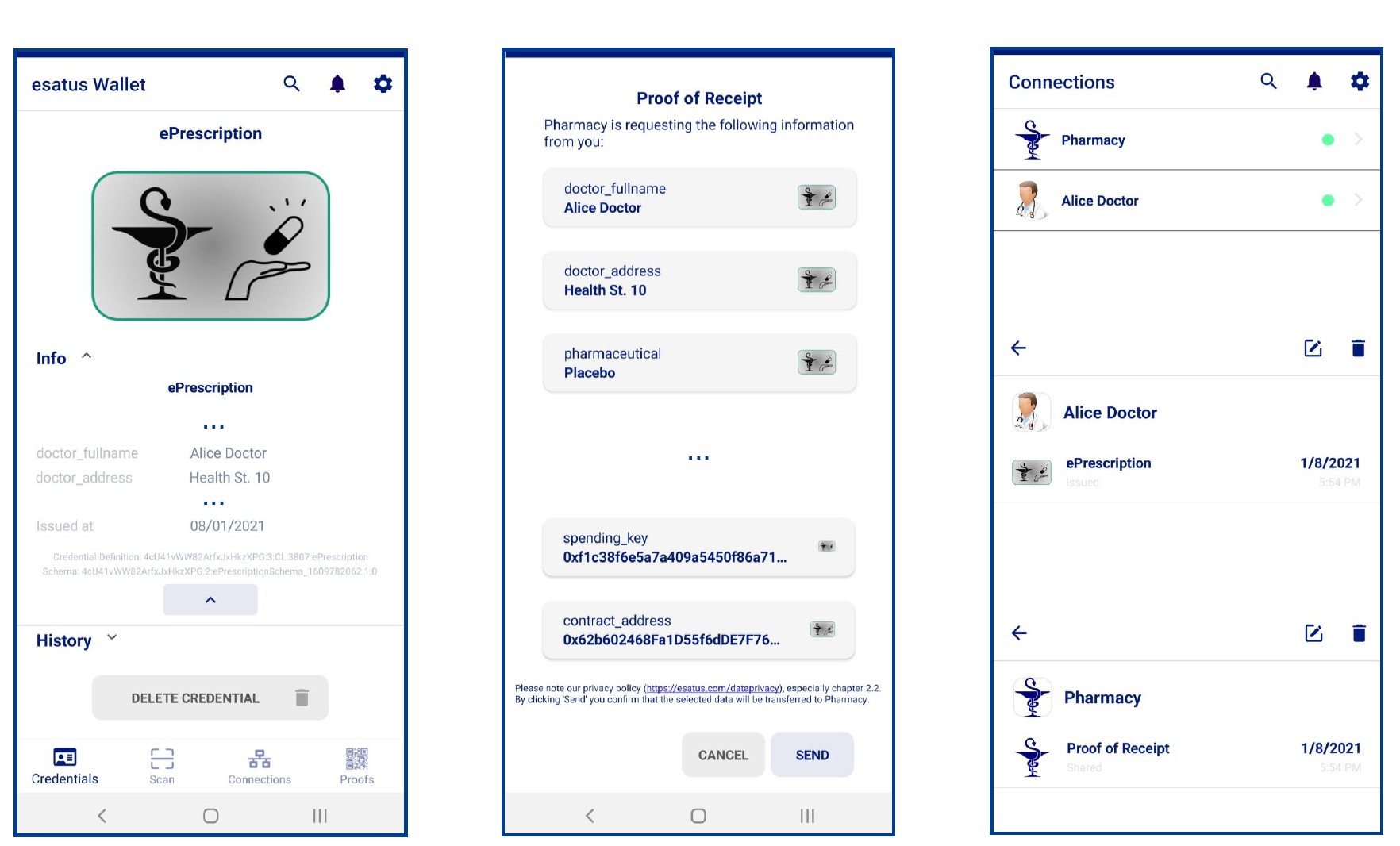}
    \caption{Screenshots from the user's wallet: Overview of the user's \acp{VC} (here: only one e-prescription for simplicity), visual check of requested attributes in a \ac{VP}, and overview of active connections as well as shared data in the esatus wallet.}
    \label{fig:screenshots}
\end{figure}

Mobile wallets can connect with a cloud agent through scanning a QR-code. Alternatively, existing communication channels, such as an e-mail, can be used for the initial exchange of information, e.\,g., in the form of a link, on how to establish a connection. The QR-code contains the cloud agent's \ac{URL} endpoint as well as further meta-information. It can be static or personalized, and it must be accessible to the patient, e.g. when checking in at the doctor, or on a health insurance's or pharmacy's website. At first, the health insurance issues an insurance \ac{VC} to the patient. Upon a doctor visit, the patient establishes a connection with the doctor's cloud agent and is subsequently asked for a presentation of their insurance \ac{VC} to map the connection to master data like the patient's name and insurance number. Next, the doctor can fill out the contents of the \ac{e-prescription} on a simple web interface or further automated programs, as they would do for any electronic and machine-readable prescription. By selecting the connection to the patient, a doctor can trigger the issuance of the \ac{e-prescription} \ac{VC} to the patient and the creation of the associated token in the doctor's smart contract. At the pharmacy's branch or website, the patient only needs to scan a static QR-code, which the wallet can again use to establish an \ac{E2E} encrypted connection to the pharmacy. It is important to understand that this connection can be unique and non-correlative for each pharmacy visit, thus allowing for privacy by avoiding metadata analysis. This action automatically triggers a proof request from the pharmacy. Consecutively, the patient must select the \ac{e-prescription} they want to spend and the respective insurance information. Upon confirmation, the patient can then share the requested attributes. During all these processes, the patient has full control over which of their personal data is shared, which connections they have established yet, and also an overview of the data that they have shared. We present screenshots from the wallet during and after the process in figure~\ref{fig:screenshots}.

%% file: 06-Evaluation.tex
\section{Evaluation}
\label{sec:evaluation}

Our proposed solution aims to address the design objectives that we determined in section~\ref{sec:related_work}. Thus, we aim at solving the issues implied by paper-based prescriptions and centralized as well as exclusively blockchain-based \acp{e-prescription}. In the following section, we evaluate our design and implementation along these design objectives.

We address multiple requirements of digital \ac{e-prescription} management systems by building on the \ac{SSI} paradigm. We rely on \ac{SSI} for the interactions of the involved stakeholders, enabling bilateral and hence decentralized data exchange. First, this approach allows users to exercise \textbf{disclosure control}, as the standardized \acp{VC} containing prescription-related information and the spending key of the associated token is transferred to patients and stored only in their digital wallet of choice. Accordingly, the physical prescription is digitized through a \ac{VC} that is stored only on the patient's device and in particular not on a centralized or decentralized ledger that would allow to track patients' interactions with doctors and pharmacies. Our design based on bilateral communication between stakeholders is, therefore, more privacy preserving than alternative solutions such as private channels in a Hyperledger Fabric permissioned blockchains. This also allows to avoid the corresponding scalability challenges of partially replicated storage of the private data’s hashes on-chain~\cite{guggenberger2022fabric}. 
The data minimization capabilities offered by existing implementations of digital wallets and especially the opportunity to leverage selective disclosure gives patients additional \textbf{privacy} and considerably more \textbf{control} over which data they share than they would have with a physical prescription.

Furthermore, the \ac{SSI} paradigm allows patients to interact bilaterally with their doctors and pharmacies. As the authenticity of \acp{VP} derived from \acp{VC} can be verified through the issuer's signature, no direct communication between doctors and pharmacies is necessary to verify the correctness of information included in the \ac{e-prescription}. This allows for \textbf{pharmacy independence} as well as \textbf{verifiability} independent of doctors. The \ac{e-prescription}'s integrity can be checked purely cryptographically, and the issuing doctor's identity and authorizations can be verified through a lookup on the public permissioned Hyperledger Indy blockchain. In contrast to centralized architectures, our solution does not rely on one \textbf{central control point aggregating patients' data} and, therefore, also avoids a single point of failure, a requirement that authors have emphasized in the past~\cite{wu2018toward}. As we rely on bilateral interactions and a highly resilient blockchain infrastructure for information that needs to be publicly accessible, we also achieve \emph{decentralization} and high \emph{availability} guarantees. All additional data required for processing \acp{e-prescription} is stored on the patient's device. Storing the tokens' spending key in the \acp{VC} also implies that keys do not have to be transferred to the user through an additional technical component. Accordingly, token-related \textbf{key management} does not involve patients directly, which decreases complexity for them throughout the prescribing and redeeming process. Instead, our architecture requires patients to use one digital wallet that they may already use in other contexts. This also addresses interoperability concerns, which are currently often solved by reliance on central authorities offering web interfaces.
Because our system provides confidentiality through bilateral interactions and disclosure control, integrity through the verifiability of the information referenced in \acp{e-prescription}, and availability, we also adhere to the general principles of information security~\cite{whitman2011principles}.     

Our demonstration uses a digital wallet and institutional agents that follow common standards for \acp{DID} and \acp{VC}; in general, we do not require dedicated \ac{e-prescription}-related functionalities in the digital wallet. Consequently, the exchange of information between patients and other stakeholders follow common standards for digital identity management and can be used also when \ac{SSI}-standards develop further in the future. Furthermore, the JSON-based structure of \acp{VC} allows to follow international standards regarding the semantics of \ac{e-prescription} systems such as directed by the~\citeauthor{EuGuidelinesePres2014}. The respective schema can be published and made available for all participating actors on the Hyperledger Indy blockchain. Thus, we argue that our architecture ensures \textbf{interoperability}. However, we also identify potential for future research with regards to this aspect. As large-scale health systems often embed \acp{e-prescription}, integration with existing systems like, e.g., electronic health records, should be tested for. Nevertheless, by avoiding lock-in effects induced by data silos and using generic components like standardized digital wallets, we argue that our architecture will likely integrate into other systems without major design changes. 

Apart from leveraging digital wallets with the \ac{SSI} paradigm to satisfy requirements of decentralized \ac{e-prescription} systems, we rely on blockchain technology to address additional requirements, ensuring \textbf{double-spending prevention} through the use of blockchain tokens. In our proposed solution, each \ac{e-prescription} \ac{VC} issued involves the creation of a corresponding blockchain token. In analogy to paper-based prescriptions, the doctor can specify the number of times the \ac{e-prescription} can be redeemed at the time of issuing the \ac{e-prescription} token, which is of particular importance for long-term medication plans.
This is due to the reason that the pharmacy redeeming the prescription will only dispense a pharmaceutical if the associated token has not already been spent. We employ separate blockchain implementations for storing \ac{SSI}-related public information and managing the tokens due to their respective properties enabling optimized service provision for the required operations. In theory, however, one blockchain implementation would be sufficient for all operations related to our design.

Furthermore, our approach achieves interoperability both on a technical as well as domain-specific level: Since only interfaces for token creation and spending are required, our approach can be adapted to other blockchains. By using openly available and tested building blocks from blockchain technology and digital identities, we avoid the need for implementing complex new functionalities. Thus, we argue that safeguarding costs arising from complexity remains a feasible objective.

Regarding the latter, we facilitate the implementation of multiple standards for \acp{e-prescription}. Thus, we adopt a flexible and, therefore, standard-agnostic approach to representing \acp{e-prescription} via JSON files~\cite{IETF2017JSON}. International document-oriented standards for health documents, such as \ac{HL7CDA}~\cite{HL72015CDA}, typically aim to facilitate the fully digital exchange and unambiguous interpretation of machine-readable documents, such as prescriptions. JSON offers these functionalities. However, JSON lacks some of the functionalities of alternatives like XML, which is used in \ac{HL7CDA}: Integrity checks are an essential part of unambiguous and reliable interpretation. We cover this aspect in our approach through the restrictions posed by the verification of a \ac{VP}: As described in section~\ref{sec:background}, the restrictions in the pharmacy's proof request ensure that a \ac{e-prescription} is created according to a schema that has previously been published and hence ensures integrity. Nevertheless, an XML-based implementation of \acp{VC} is conceptually possible. Consequently, while our approach does not implement any specific existing standard, we can address all requirements underlying common standards. Our research does not target a specific geography or regulation, but mainly aims to demonstrate the potential of decentralized technologies for providing high levels of privacy and convenience in e-prescriptions.

To ensure the practicability of our system, aspects regarding \textbf{scalability and performance} must be considered, including throughput, latency, and costs. Besides, we also assess performance metrics like the end-to-end duration of the \ac{e-prescription} creation and spending process for our prototype. Regarding costs, several observations of the system can be made. If the smart contract described in figure~\ref{fig:ePrescription_Contract} was deployed on the public Ethereum blockchain, creating the token would currently cost around 60\,USD, and the redemption of an \ac{e-prescription} around 8\,USD\footnote{The Gas costs for creating a prescription in the public Ethereum network are approximately 85,000 Gas. 1 Gas currently costs around $10^{11}$\,Wei. One Ether corresponds to $10^{18}$\,Wei, i.e., 1\,Gas corresponds to $10^{-7}$ Ether. One Ether currently costs more than 4,000~USD, so 
$$c_{\mathrm{create}}\approx\frac{4,000\,\mathrm{USD}\,\times\,1.5\,\cdot 10^{5}}{10^7}\approx60\,\mathrm{USD}.$$
Gas for spending a prescription again depends on the amount to around 14,000~Gas, i.e., around 10\,USD -- cheaper, but still way too expensive for our use case.}. Moreover, the throughput would currently be limited to around 10\,tx/s, and the latency for creating and sending the token would be in the order of a minute and highly volatile. For this reason, we decided to use a permissioned blockchain in our implementation. In the near future, sharding and second layer solutions such as optimistic and zk-rollups~\cite{gudgeon2019sok,schaffner2021scaling} could enable implementing this architecture also on public permissionless blockchains with acceptable costs.

\begin{figure}[!tb]
    \centering
    \includegraphics[width=0.55\linewidth]{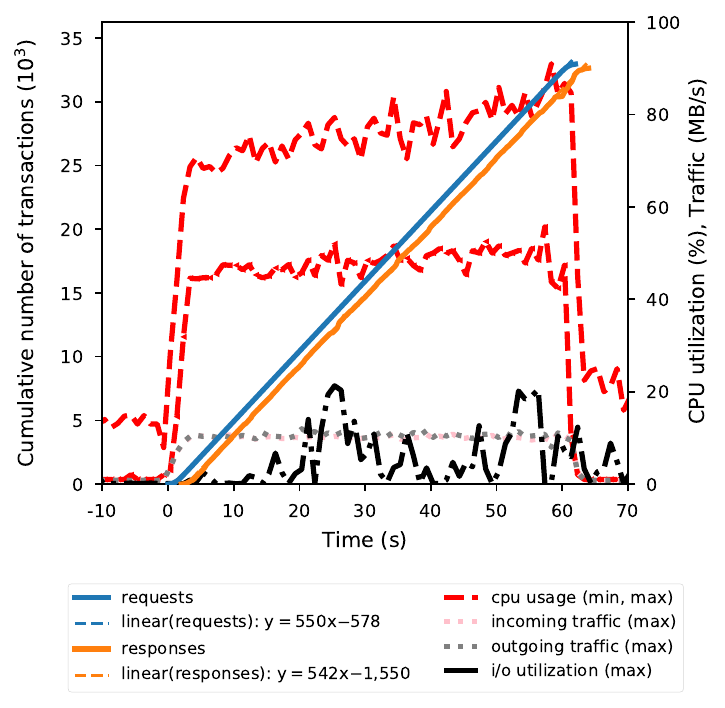}
    \caption{Benchmark of creating \ac{e-prescription} tokens on a 32-node cross-European Quorum network on \ac{AWS} m5.2xlarge instances (8\,vCPUs, 16\,GB RAM) network with \ac{IBFT} consensus at a request rate of 550\,tx/s. Max respectively min means the maximum measured over all nodes; for CPU usage, additionally max respectively min is taken over all of a node's cores first.}
    \label{fig:benchmark}
\end{figure}

To estimate whether the throughput of existing permissioned blockchains is sufficient for a hypothetical large-scale deployment of our \ac{e-prescription} management system, we conducted a performance analysis with the \ac{DLPS}~\cite{sedlmeir2021benchmarking}. The respective testing framework is available open source\footnote{Available from: https://www.github.com/DLPS-Framework/}. To simulate a cross-European \ac{e-prescription} system, we deployed a Quorum network in \ac{AWS}, with 8 nodes each in data centers in Frankfurt, Dublin, Milan and Stockholm. As mentioned above, we chose \ac{IBFT} as consensus mechanism because it is more fault-tolerant (and, thus, has a reduced performance compared to RAFT). This makes our network comparable to the design currently deployed by \ac{EBSI}, where every member state of the European Union is supposed to run a private Ethereum node\footnote{Although \ac{EBSI} does not run Quorum nodes but Hyperledger BESU instead, Quorum and BESU are at the core both based on very similar protocols~\cite{EBSIArchitecture}. Also, regarding their consensus mechanisms, they have similarities: BESU runs \ac{IBFT}2, an upgrade of the \ac{IBFT} consensus mechanism in our Quorum network.}. For these 32~nodes, we used instances from the \ac{AWS} EC2 m5.2xlarge series that have 8~vCPUs and 16\,GB RAM. This specification is on the lower bound of what \ac{EBSI} lists as requirements for running a node. With the code of the e-prescription implementation, we also publish a configuration file that fully describes the settings that we used for the benchmarking process. This configuration file can be used to reproduce our results.

We concentrated our benchmarking efforts on the \textit{createPrescription} method, as the gas costs indicate that \textit{spendPrescription} is less computationally expensive. Using the \ac{DLPS}, we found a maximum sustainable throughput of around 630\,tx/s with a latency of 2.0\,$\pm$\,0.3\,s in the described setting (see figure~\ref{fig:benchmark}). 

4.5~billion prescriptions a year get filled inside the U.S.~\cite{Mattke:2019:BlockchainPharma} and 1.5~million prescriptions per day in the UK~\cite{aldughayfiq2020digital}. The maximum of these approximations amounts to 14 prescriptions per citizen and year on average. In the \ac{EU}, this would therefore hypothetically amount to approximately 6.1~billion prescriptions per year, so an average throughput of 200\,tx/s for each creating and spending \ac{e-prescription} token is required. We assume that prescriptions will be created and spent mainly on working days between 08:00 and 17:00 and there also will be fluctuations. Thus, a multiplier of at least 3, more likely 5--10, should be taken into account, which means that around 2,000\,tx/s are required. Note that a larger throughput than 550\,tx/s can be achieved with smaller network sizes, better hardware, or using RAFT instead of \ac{IBFT}. Thus, we conclude that a large-scale implementation of our proposed solution can be considered realistic with existing technology. We also expect performance improvements caused by optimizations, improvements in hardware, and the opportunity to set up two or three separate blockchain networks for token management if performance is not sufficient by the time that all prescriptions in Europe have been digitized. Concerning costs, even running~10 of the the described Quorum blockchains would cost only around 1,000,000 USD per year for server provision on \ac{AWS}~\cite{AWS2021pricing}, which amounts to less than 10$^{-3}$\,USD per \ac{e-prescription}.

The latency for the \textit{createToken} and \textit{spendToken} methods that the doctor respectively the pharmacy need to invoke during issuing respectively verifying an \ac{e-prescription} in our implementation each take around 2.5\,s at low throughput when the block-time of \ac{IBFT} is configured to 1~second. Our measurements also yield that at higher throughput, latency does not increase considerably.

The steps involved in an \ac{SSI}-based interaction (establishing a connection, issuing a \ac{VC} or verifying a \ac{VP}) as described do not provide a significant challenge regarding scalability as the interactions are conducted bilaterally: We found that a single cloud agent that the doctor and the pharmacy run can issue and verify at least two \acp{VC} per second, which should be sufficient even for a medical office with several doctors. Moreover, these clients can be scaled horizontally if necessary. The processes conducted on the Hyperledger Indy network do not negatively affect scalability: Doctors can update their revocation registries in batches, e.\,g., once a day; and revocation is likely to happen only rarely anyway. The creation of credential schemas or credential definitions only has to be conducted during onboarding processes of doctors and, thus, does not require high throughput. Finally, a single Indy node can serve several hundred read operations per second\footnote{We measured this on our own with the same method as applied for the Quorum benchmark above.}, so a medium-sized Indy network with a two-digit number of nodes can handle sufficiently many read requests per second. Note that if no proof of revocation is necessary and a pharmacy stores the \acp{DID} and public keys (credential definitions) of the doctors in their vicinity and only updates this local copy occasionally, a \ac{VP} would not even need a single write or read operation on the Indy blockchain. We conclude that our \ac{e-prescription} management system provides sufficient scalability and performance for hundreds of millions of users. 

Regarding the patient-side processes, it takes a total of around 15~seconds for the patient to open their wallet, establish a connection with the doctor, and accept that the \ac{e-prescription} is issued to and stored in the patient's smartphone wallet (this involves two user-side confirmations). A similar amount of time is needed from the patient opening their wallet, scanning the QR code for the redemption process, until the pharmacy has verified the \ac{VP} and that the token has not been spent yet. We consider this a reasonable time for the end-to-end automation of the paper-based process, although practical trials would ultimately need to confirm patient's satisfaction with the process' speed. 

Privacy and security become increasingly important in networked systems and related threats to blockchain-based systems are well-known \cite{schlatt2022attacking}. To formally assess the security and privacy implications of the developed system, we performed a thorough security analysis following the STRIDE framework~\cite{howard2006security}. This framework assesses the security of a system by evaluating threats grouped in six categories. These categories comprise spoofing, tampering, repudiation, information disclosure, denial of service, as well as escalation of privileges. Thus, it covers several security properties including authorization, confidentiality, integrity, and availability~\cite{khan2017stride}, which are also reflected in our requirements. The respective evaluation can be performed both component-based as well as interaction-based to provide a granular analysis of the system~\cite{shostack2014threat}. We assume that the open-source components that we build on  were already analyzed for their security properties. Consequently, we opt for an interaction-based analysis to assess the interplay of the different components making up our system and, thus, our contribution. We consider two main interactions relevant in our use case: \textit{(I1)} the doctor issues a prescription to the patient and \textit{(I2)} the patient redeems a prescription in a pharmacy. To make our considerations visible, we first documented the interactions through detailed sequence diagrams (see Figures~\ref{fig:seqDiagram_Issuance_I}~and~\ref{fig:seqDiagram_Verification}). Table~\ref{tab:stride} features our analysis' results regarding their respective susceptibility to STRIDE threats.

\renewcommand{\arraystretch}{1.5}
\begin{table}[h]
    \centering
    \vspace{0.25cm}
    \begin{tabular}{c|c|c|c|c|c|c}
        \toprule
        Interaction & S & T & R & I & D & E \\
        \midrule
        \textit{I1} & x & o & o & x & o & x \\
        \textit{I2} & x & o & o & ? & o & o \\
        \bottomrule
    \end{tabular} 
    \vspace{0cm}
    \caption{Results of the STRIDE threat analysis.}
    \label{tab:stride}
    \end{table}

We identified four major threats (indicated through x~and~?) to security and privacy in our proposed system, which we briefly elaborate on. First, we identified that man-in-the-middle attacks related to spoofing are potentially possible in the current system design. In practice, setting up interactions between parties involves the patient scanning a QR-code. By replacing the correct QR code, which is used to start a connection with the doctor or, respectively, the pharmacy, an attacker could pretend to be the intended recipient and subsequently engage in an interaction with the patient. In this context, the attacker could, for example, retrieve information contained in the patient's \acp{VC}. Thus, this attack represents a spoofing threat. Because the attack can be conducted in the same manner in both interaction scenarios, we count both instances as one. However, there are several means to identify the relying party. For example, the process may require an authenticated out-of-band channel for transmitting the QR-code to initiate the e-prescription issuance, and regular audits of the QR-codes displayed in pharmacies. Furthermore, we can achieve this through a lookup in a DLT-based trust registry. Alternatively, the verification of established SSL~or~\ac{QWAC} certificates is possible. The implementation of these approaches is currently in progress, for instance, in the Lissi SSI-wallet. Second, the current setup discloses information about the number of prescriptions issued by each individual doctor to all Quorum nodes. While this makes already much less information visible on-chain than the related work that we presented in section~\ref{subsec:litrev}, it is a shortcoming that future work should address, e.g., through \acp{ZKP}. We discuss this in more detail in the conclusion (section~\ref{sec:conclusion}. Corresponding confidentiality issues could also be mitigated through the suitable restriction of participating nodes; the extreme solution being a centralized solution for double-spending prevention that does not harm the protection of sensitive data by design. Third, we found that doctors may prescribe arbitrary prescriptions in the current design, resulting in an elevation of privileges. This shortcoming can be addressed by adapting the proposed e-prescriptions smart contracts. In specific, regulatory bodies creating the smart contracts for the doctor’s public key can include public information about which medicals the doctor may or may not prescribe. A more privacy-oriented alternative would be to issue e-prescriptions as chained credentials, i.e., doctors issue e-prescriptions together with a digital certificate that a regulatory body issued to them. Consequently, pharmacies could check that the doctor that issued the e-prescription was him-/herself entitled to do so by the doctor’s certification. While this additional mechanism is feasible, it lies outside the scope of the purpose of our prototype. 
Fourth, as self-sovereign identities represent a novel concept, it is still contested whether information can be deduced from revocation registries using cryptographic accumulator. However, an intermediate assessment from large-scale pilots in Germany suggests that relying on revocation registries using cryptographic accumulators is indeed compliant with the \ac{GDPR}.

We discuss some of these important guidelines and the corresponding opportunities for further research in more detail in the following two chapters. Nonetheless, we can summarize that we can address of the remaining potential security threats and, therefore, claim that our approach is secure.

%% file: 07-Discussion.tex
\section{Discussion}
\label{sec:discussion}

The following section offers a discussion of four important implications related to the contribution of our design.
First, the role and necessity of applying blockchain in the context of \ac{SSI} systems remain debatable. In our design, we deliberately chose to use a blockchain for several distinct purposes. As a transparent, highly available, and tamper-resistant decentralized infrastructure, one blockchain implementation serves as a permanent ledger storing information of public interest in the context of \ac{SSI}: profiles of issuing doctors, privacy-preserving revocation registries for \acp{e-prescription}, and schemas thereof. These functionalities are relevant in the context of any \ac{SSI} system, which is why we deem it reasonable to include a blockchain for these purposes in the respective designs. Nevertheless, these functionalities are also achievable using centralized infrastructures, however giving up some of the advantages coming with a blockchain such as independence from \acp{CA} (in the context of \acp{e-prescription}) and availability. While in many \ac{SSI} applications the re-use of \acp{VC} is desired, such as for national ID \acp{VC}, for \acp{e-prescription} avoiding possibilities for double-spending is of utmost importance regarding \acp{e-prescription}. As a result, we incorporated a second blockchain implementation managing unique tokens corresponding to each \ac{e-prescription} \ac{VC}. In principle, a centralized system could also be used for this from a pure privacy perspective, as the data required for the double-spending check is no more confidential in our construction. We suggest a design that separates the double-spending check from the layer of exchange of confidential data for any \ac{SSI} use case requiring a restricted number of usages in \acp{VC}. Separating the restriction regarding double-spending (token) from invalidation (revocation) also allows to prevent multiple redemptions but at the same time reusing the \ac{VC} for information exchange, e.\,g., to check for cross-reactions of medicals, without a limit.

Second, it is important to note that \ac{SSI} and blockchain technologies are characterized by open-source software philosophies and are often freely available. Moreover, the respective technologies do not include a centralized, controlling party by design. Therefore, the stakeholders need to establish adequate governance structures to ensure the creation of a functioning ecosystem, including secure processes outside of the technical system proposed~\citep{schlatt2021designing}. In the context of prescriptions, the central entities in the respective communities could take up important roles in this regard. For example, stakeholders must agree on mechanisms for certifying doctors or the attributes including their data types specified in \ac{e-prescription} \acp{VC}. In this paper, we focus on the technical design, since there is valuable related work on governance, both in the realm of blockchain technology~\cite{beck2018blockchain} and \ac{SSI}~\cite{joosten2021ssigovernance}.

Third, prior research indicates that the acceptance and usefulness of \acp{e-prescription} increases when the required functionalities are embedded in a larger ecosystem~\cite{tamblyn2006development}. This could, for example, include generic systems for managing and exchanging patient health data. To this end, the generic nature of the components used in \ac{SSI} systems can provide several promising opportunities~\cite{Guggenberger2022Accept}. Network effects can occur through the interplay of a \ac{VC}-based \ac{e-prescription} with other \acp{VC} such as digital insurance cards, ID cards, or vaccination credentials. Thereby, the mutual utility of such \ac{VC} can be improved, making the \ac{e-prescription} framework even more versatile, e.\,g. for a proof of identity in telemedicine, communicating with an insurance company, or digitizing the reimbursement process for prescriptions. 

Fourth, by combining \ac{SSI} and blockchain, we suppose that our solution addresses several regulatory requirements, especially those imposed by the strict EU \ac{GDPR}. 
These include principles for processing personal data such as data minimization, integrity, and confidentiality, as well as lawfulness, fairness, and transparency (Article 5 \ac{GDPR})~\cite{GDPR2016}. We address data minimization and confidentially by relying on \ac{ZKP} through using selective disclosure and releasing only the data requested by the pharmacy. No third parties have access to this data. This is complemented by the use of unique \acp{DID} for each connection as well as \ac{E2E} encrypted connections to provide a private connection between the parties involved and to minimize the risk of unintended correlation~\citep{schlatt2021designing}.
Nevertheless, \acp{ZKP} have been rarely used for signature schemas in practice so far and are thus still uncharted territory from a regulatory perspective. We  respect the rights of data subjects, such as the right to erasure (``right to be forgotten'') (Article~17~\ac{GDPR}), as no patient-related data or metadata is stored on the ledger. Rather, by relying on the \ac{SSI} paradigm, we give the user the responsibility for their own data such that the right to erasure and the right to rectification are also adhered to (Article~16~\ac{GDPR}). This design choice also addresses the right to data portability (Article~20~\ac{GDPR}) as we use components oriented at the \ac{DID} and \ac{VC} standards and interoperable wallets for implementing the paradigm.

Yet, owing to the scope of our prototype, several domain-specific requirements remain to be implemented through evaluation of our system in practice. For instance, the current prototype allows any doctor to prescribe any medicine in theory. As remedies for each of these shortcomings exist, we propose future research to address those. 

Our system relates to and addresses shortcomings of related research. Compared to existing work identified in our literature review (Section \ref{sec:method}), we facilitate data privacy in blockchain-based \ac{e-prescription} management systems by avoiding storage of personal information on-chain, which has been identified as a limitation in prior research~\citep{thatcher2018pharmaceutical}. Furthermore, we allow for bilateral exchange of cryptographically verifiable \acp{e-prescription} rather than simply allowing for data integrity checks through blockchain, like~\citet{li2019dmms} and~\citet{meena2019preserving}'s solution. In addition, we extend \ac{SSI}-based solutions by offering a double-spending prevention through blockchain-based token. Existing work by~\citet{cadoret2020proposed} as well as~\citet{gropper2016powering} exhibit these shortcomings. Thus, we combine and extend prior research through our system for \ac{e-prescription} management.

The design process and evaluation of our solution allowed us to derive design principles for decentralized information systems that involve the exchange of sensitive data and a control on the number of usages. These design principles elevate our research contribution for theoretical discussion, and the additional level of abstraction allows practitioners and researchers to apply our observation in scenarios with similar challenges~\citep{hevner2004design,gregor2013positioning}. As our evaluation suggests, \ac{SSI} can avoid data silos, provide a standardized interface for the exchange of verifiable information whereas blockchains can solve the challenge of transferring value in a decentralized system. In contrast, blockchains cannot be used for the exchange of verifiable, sensitive information due to their inherent replication and immutability; yet, they allow for control on the number of usages of \acp{VC} across different bilateral interactions. Consequently, we formulate the corresponding design principles as follows:
\begin{itemize}
    \item[1.] Use verifiable credentials stored in a digital wallet to provide sensitive and verifiable user information to services.
    \item[2.] Implement vouchers through creating a token and including its spending secret to the digital certificate. This benefits usability and ease of implementation because users do not require a mobile app beyond their digital wallet. 
\end{itemize}
Furthermore, several practitioners noted our approach's potential of increasing efficiency and reducing costs. This is specifically so when different stakeholders can refer to the same software (like issuing and verifying components) and credential ecosystems (consensus on the list of trusted issuers and the semantic interpretation of attributes referenced in \acp{VC}). For example, adding the health insurance card allows for the fast onboarding of a patient at a doctor or the issuance of a receipt at the pharmacy. Patients can directly present the latter to their health insurance for reimbursement in combination with their health insurance \ac{VC}.
Thus, we formulate this in a third design principle:
\begin{itemize}
    \item[3.] Create additional value by building an ecosystem in which \acp{VC} can be combined and used repeatedly in different contexts.
\end{itemize}

%% file: 08-Conclusion.tex
\section{Conclusion}
\label{sec:conclusion}

To solve the challenges of harmonizing sensitive data exchange and double-spending prevention, we study the case of \acp{e-prescription} and design and implement a decentralized \ac{e-prescription} management system using blockchain technology and \ac{SSI} digital wallets. We evaluate the proposed system along requirements derived from a literature review. Thus, we contribute a generic design as well as an illustrative implementation of an \ac{e-prescription} management system. The findings deduced from our evaluation offer insights into the opportunities and remaining challenges regarding the development of decentralized systems dealing with sensitive data and business-logic that involves multiple stakeholders. Our design and implementation also aid practitioners in developing systems beyond the health sector with similar requirements. We conclude the paper by identifying our work's limitations as well as highlighting opportunities for future applications and research.

During implementation, we experienced technical limitations, which allow us to derive several promising avenues for future research. First, to date credential delegation mechanisms that allows a patient to forward their \ac{e-prescription} \ac{VC} (including the spending key) to another user as proposed by~\citet{camenisch2017delegation} are not integrated in common \ac{SSI} frameworks. This feature is desirable as proxies are of particular importance in the healthcare sector, e.g., in the case of old age or severe illness. The missing functionality of credential delegation also implies that certificate chains are currently not supported in our prototype. On the one hand, credential chains would allow to limit the type of medicines that a doctor may prescribe and thus solve on of the possible threats to our system (see thread E - elevation of privileges). As such, regulatory bodies could issue a \ac{VC} to doctors authorizing the issuance of certain medicine. In turn, doctors could then issue the~\ac{e-prescription} \ac{VC} accompanied by a delegated version of their \ac{VC}, thus proving their authority to prescribe the corresponding medical. On the other hand, credential delegation would simplify the delegation of tasks in a surgery or to fetch a medical on a pharmacy on behalf of the recipient. For instance, if a doctor wanted to delegate the issuance of \ac{e-prescription} \acp{VC} to their employees, these would currently need their own public \ac{DID} and credential definition to be individually recognized by pharmacies as trusted issuers of \acp{e-prescription}, and their own prescription smart contract. The support of certificate chains and delegation in \ac{SSI} could make such a system substantially easier to implement, govern, and scale. Furthermore, complexity may be reduced by reproducing the proposed concept with only one blockchain as soon as the respective tools and frameworks are available. 

Second, as our security analysis showed, further investigations are required on whether the revocation registry can be stored on a blockchain from a regulatory perspective or whether it should be held in central databases. While we aimed to comply with the current \ac{GDPR} policies through the cryptographic accumulator-based design of revocation registries, we suggest a more thorough evaluation with legal experts and extending the regulations under considerations to \ac{CCPA} or \ac{HIPAA}.

Third, while patients' privacy was of utmost priority for the architectural design, we identified potential for improvement with regards to the tracking of doctors' activities through our threat analysis (see thread I - information disclosure). In specific, the implementation currently relies on the use of long-term key pairs by doctors for deploying smart contracts. Thus, as we highlighted in our security analysis, blockchain nodes can see how many prescriptions a doctor has issued in our current implementation. This shortcoming could be addressed through the usage of \ac{ZKP} in future work, which are used by, e.g., cryptocurrencies such as Z-Cash~\citep{sasson2014zerocash}. Thus, we propose introducing a single pool for creating and spending \ac{e-prescription} tokens and replacing the currently used anonymized valid and redeemed \ac{e-prescription} tokens by cryptographic commitments and nullifiers to prevent the tracking of doctors' activities and, thus, minimize remaining risks regarding information disclosure. In addition, the architecture could be extended to add the respective e-prescription to the patient's health record, or to make it accessible to other domain-specific applications that, e.g., check whether there are potentially interactions between different medicals. For this purpose, a receipt could be given to the patient as a \ac{VC}, which they can then present to their insurance or application maintaining the respective health records. Alternatively, this could be done directly by the doctor or pharmacy in a bilateral interaction, where the doctor or pharmacy forwards the information contained in the e-prescription \ac{VC} they issued, respectively the VP they received.To reduce the spoofing risks identified in our threat analysis (see thread S - spoofing, several approaches are currently being implemented, such as identifying the party to which a connection is established through their on-chain identifier and public key, through an initial \ac{VP} in which they reveal a digital certificate that they received from a certifying body (e.g., an SSL certificate or a \ac{QWAC} established in the European eIDAS framework)~\citep{schellinger2022mythbusting}.

Fourth, we have not yet tested our prototype in practice. As the system's potential users and legacy systems are highly heterogeneous, a field test and subsequent evaluation with end users remains a significant limitation and will be one of the next steps in our research. As such, also the costs for implementing our system as proposed cannot be quantified at this stage of our research and remains a promising avenue for further research. However, when we presented our prototype to practitioners with expertise on blockchain and \ac{SSI}, their feedback gave us confidence that the proposed architecture can experience acceptance by different stakeholders. Yet, an extensive application in practice potentially offers more insights on integration with common legacy systems and other aspects, which can hardly be simulated in research settings.

The health sector is currently undergoing rapid changes due to the constantly advancing digitalization of society. Health data is highly sensitive and therefore requires the application of secure and privacy-preserving technologies. Owing to their respective properties in those regards, the combination of blockchain technology and \ac{SSI} offers promising opportunities for the digital transformation not only of the health care sector. The need for a high degree of user control on confidential data is also present in many other scenarios, such as interactions of citizens with their government, or business-to-business interactions. We recommend to implement and evaluate similar designs in other use cases in practice requiring both privacy as well as double-spending protection. Due to the novelty of the applied technologies in our proposed system, studies about the interactions of users with respective systems are still rare. Thus, we encourage researchers and practitioners alike to take up our work and further study the respective interactions and resulting implications as well as improvements for the implementation of our design.

%% file: 91-Appendix.tex
\begin{figure}[H]
\begin{lstlisting}[basicstyle=\footnotesize\ttfamily]
pragma solidity ^0.5.16;

contract PrescriptionContract {
    address admin;
    
    struct Prescription {
        address issuer; //the doctor's public key
        uint remainingRedemptions; //number of times that the prescription can still be spent    
    }
    
    constructor() public {
        admin = msg.sender;
    }
    
    modifier onlyAdmin() {
        require (msg.sender == admin, "Sender is not the admin of the contract");
        _;
    }
    
    // a prescription token registry, associating public keys with prescriptions
    mapping (address => Prescription) public prescriptions;
    
    function create(address patient, uint number) public onlyAdmin() {
        prescriptions[patient] = Prescription({issuer: msg.sender, remainingRedemptions: number});
    }
    
    function spend() public {
        require (prescriptions[msg.sender].remainingRedemptions >= 1, 
            "Already spent");
        prescriptions[msg.sender].remainingRedemptions--;
    }
}
\end{lstlisting}
\caption{The e-prescription smart contract implementation in Solidity that each doctor deploys for the token management of the e-prescriptions that they issue.}
\label{fig:ePrescription_Contract}
\end{figure}

\clearpage
\begin{figure}[H]
    \centering
    \includegraphics[width=0.8\linewidth, trim=0cm 1cm 0cm 0cm, clip, page=5]{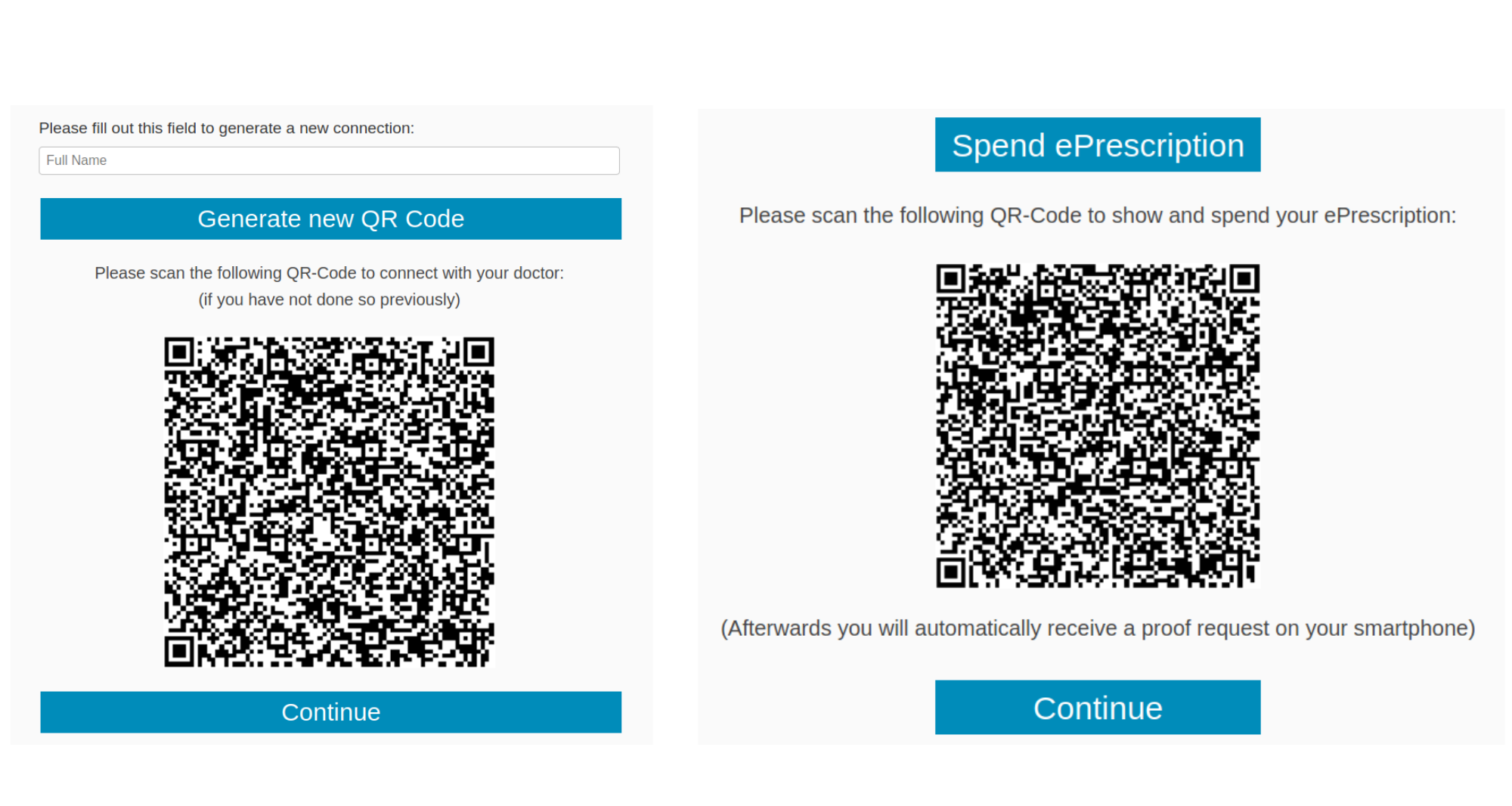}
    \caption{Frontend for establishing a connection between the user's smartphone and the doctor resp. the pharmacy.}
    \label{fig:frontend}
\end{figure}

\clearpage
\begin{figure}[H]
    \centering
    \includegraphics[width=0.8\linewidth, trim=0cm 26cm 0cm 0cm, clip]{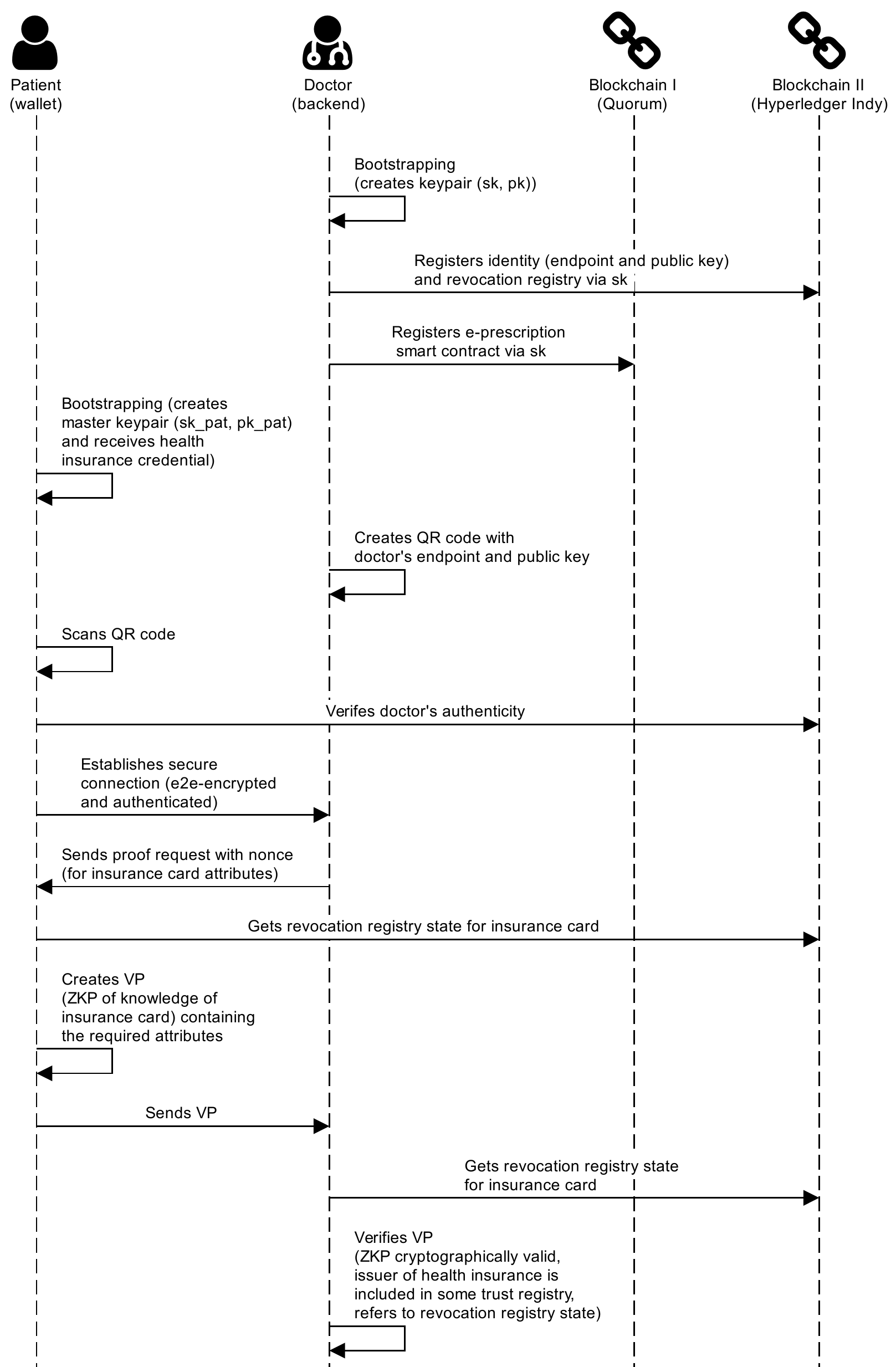}
    \caption{\ac{e-prescription} issuance.}
    \label{fig:seqDiagram_Issuance_I}
\end{figure}

\clearpage

\begin{figure}[H]
    \ContinuedFloat
    \centering
    \includegraphics[width=0.8\linewidth, trim=0cm 0cm 0cm 41.5cm, clip]{Figures/Issuance.pdf}
    \caption[]{\ac{e-prescription} issuance (cont.)}
    \label{fig:seqDiagram_Issuance_II}
\end{figure}

\clearpage

\begin{figure}[H]
    \centering
    \includegraphics[width=0.8\linewidth]{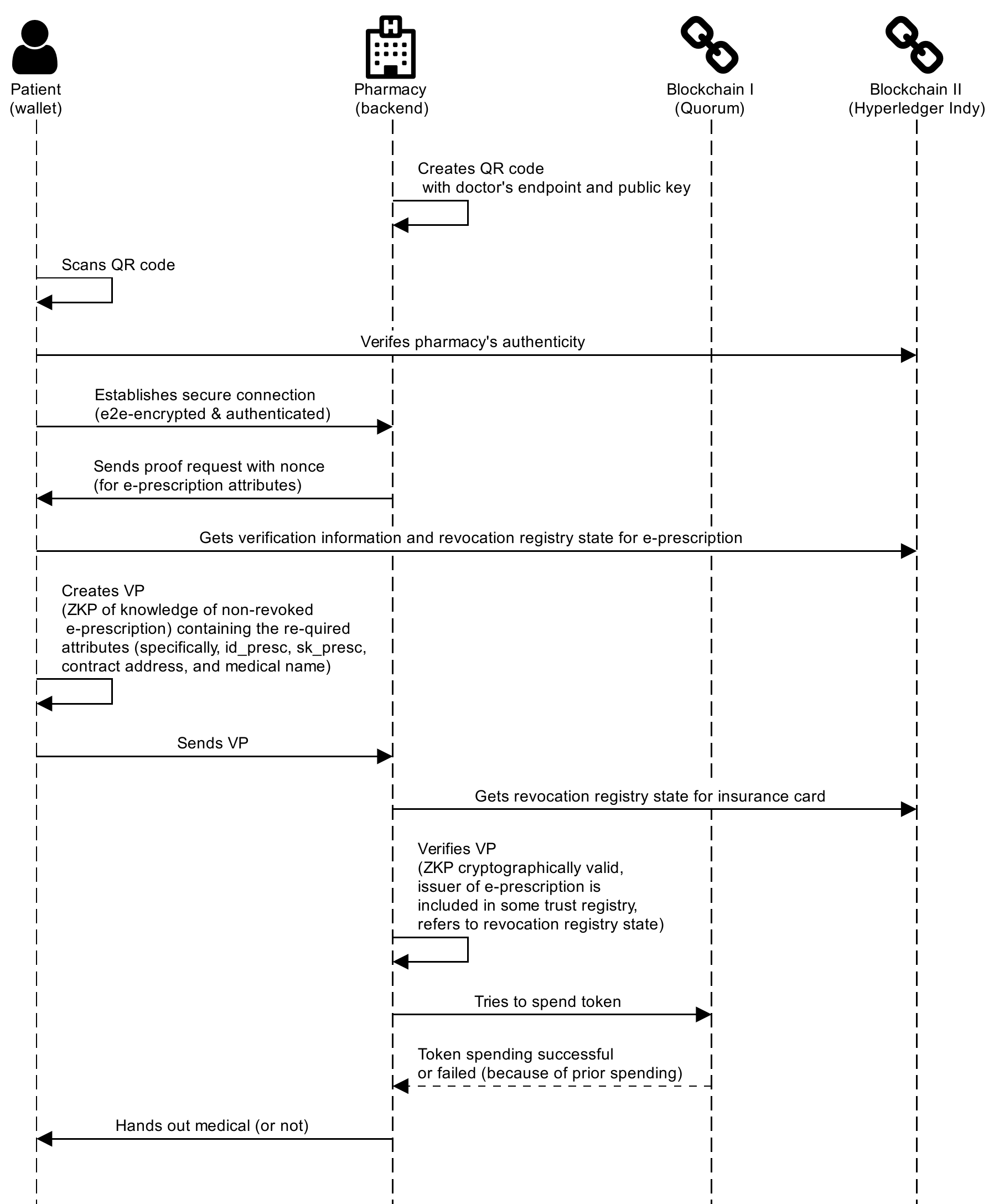}
    \caption{\ac{e-prescription} redemption.}
    \label{fig:seqDiagram_Verification}
\end{figure}